\documentclass[12pt]{article}
\usepackage{etex}
\usepackage[hidelinks]{hyperref}
\usepackage{fontenc}
\usepackage[latin1]{inputenc}
\usepackage{amsmath}
\usepackage{amsfonts}
\usepackage{amssymb}
\usepackage[affil-it]{authblk}
\usepackage{amsthm}
\usepackage{fancyhdr}
\usepackage{enumerate}
\usepackage{mathtools}
\usepackage{mathdots}
\usepackage{tikz}
\usepackage{enumitem, color, amssymb}
\usepackage{graphicx}
\usepackage{makeidx}
\usepackage [left=2cm,top=2cm,right=2cm, headheight=15pt]{geometry}
\usepackage{color}
\usepackage{caption}
\usepackage{float}
\restylefloat{table}
\usepackage{bbm}
\usepackage{bm}
\usepackage{xcolor}
\usepackage{tikz}
\usepackage{multirow}
\usetikzlibrary{shapes,decorations,arrows,calc,arrows.meta,fit,positioning}
\tikzset{
    -Latex,auto,node distance =1 cm and 1 cm,semithick,
    state/.style ={circle, draw, minimum width = 1 cm},
    point/.style = {circle, draw, inner sep=0.04cm,fill,node contents={}},
    bidirected/.style={Latex-Latex,dashed},
    el/.style = {inner sep=2pt, align=left, sloped}
}
\makeatletter
\def\squarecorner#1{
    \pgf@x=\the\wd\pgfnodeparttextbox%
    \pgfmathsetlength\pgf@xc{\pgfkeysvalueof{/pgf/inner xsep}}%
    \advance\pgf@x by 2\pgf@xc%
    \pgfmathsetlength\pgf@xb{\pgfkeysvalueof{/pgf/minimum width}}%
    \ifdim\pgf@x<\pgf@xb%
        \pgf@x=\pgf@xb%
    \fi%
    \pgf@y=\ht\pgfnodeparttextbox%
    \advance\pgf@y by\dp\pgfnodeparttextbox%
    \pgfmathsetlength\pgf@yc{\pgfkeysvalueof{/pgf/inner ysep}}%
    \advance\pgf@y by 2\pgf@yc%
    \pgfmathsetlength\pgf@yb{\pgfkeysvalueof{/pgf/minimum height}}%
    \ifdim\pgf@y<\pgf@yb%
        \pgf@y=\pgf@yb%
    \fi%
    \ifdim\pgf@x<\pgf@y%
        \pgf@x=\pgf@y%
    \else
        \pgf@y=\pgf@x%
    \fi
    \pgf@x=#1.5\pgf@x%
    \advance\pgf@x by.5\wd\pgfnodeparttextbox%
    \pgfmathsetlength\pgf@xa{\pgfkeysvalueof{/pgf/outer xsep}}%
    \advance\pgf@x by#1\pgf@xa%
    \pgf@y=#1.5\pgf@y%
    \advance\pgf@y by-.5\dp\pgfnodeparttextbox%
    \advance\pgf@y by.5\ht\pgfnodeparttextbox%
    \pgfmathsetlength\pgf@ya{\pgfkeysvalueof{/pgf/outer ysep}}%
    \advance\pgf@y by#1\pgf@ya%
}
\makeatother

\pgfdeclareshape{square}{
    \savedanchor\northeast{\squarecorner{}}
    \savedanchor\southwest{\squarecorner{-}}

    \foreach \x in {east,west} \foreach \y in {north,mid,base,south} {
        \inheritanchor[from=rectangle]{\y\space\x}
    }
    \foreach \x in {east,west,north,mid,base,south,center,text} {
        \inheritanchor[from=rectangle]{\x}
    }
    \inheritanchorborder[from=rectangle]
    \inheritbackgroundpath[from=rectangle]
}
\usepackage{titlesec}
\usepackage[linesnumbered,ruled]{algorithm2e}
\usepackage{fourier} 
\usepackage{array}
\usepackage{makecell}
\usepackage{subcaption}
\usepackage[square,numbers]{natbib}
\setcitestyle{authoryear,open={(},close={)}}

\titleformat*{\section}{\LARGE\fontfamily{ptm}\selectfont}
\titleformat*{\subsection}{\Large\fontfamily{ptm}\selectfont}

\newcommand*\circled[1]{\tikz[baseline=(char.base)]{
            \node[shape=circle,draw,inner sep=1pt] (char) {#1};}}

\title{ 
Causal predictive inference and target trial emulation }
\author{Andrew Yiu$\hspace{0.3mm}^{*,1}$, Edwin Fong$\hspace{0.3mm}^{2}$, Stephen Walker$\hspace{0.3mm}^{3}$, Chris Holmes$\hspace{0.3mm}^{1,4}$}
\date{\today}

\newtheoremstyle{mytheoremstyle} 
        {\topsep}                    
        {\topsep}                    
        {\itshape\fontfamily{ptm}\selectfont}                   
        {}                           
        {\fontfamily{ptm}\selectfont\scshape}                   
        {:}                          
        {.5em}                       
        {}  
\theoremstyle{mytheoremstyle}
\newtheorem{example}{Example}
\newtheorem{theorem}{Theorem}
\newtheorem{proposition}{Proposition}
\newtheorem{corollary}{Corollary}

\newtheorem{assumption}{Assumption}

\newtheorem*{remark*}{Remark}

\begin{document}

{\fontfamily{ptm}\selectfont

\maketitle

\newcommand{\iid}{\stackrel{iid}{\sim}}
\newcommand{\fnhat}{\hat{F}_{n}}
\newcommand{\fhhat}{\hat{f}_{h}}
\newcommand{\gnhat}{\hat{G}_{n}}
\newcommand{\thehat}{\hat{\theta}_{n}}
\newcommand{\sumi}{\sum\limits_{i=1}^n}
\newcommand{\sumj}{\sum\limits_{j=1}^N}
\newcommand{\vars}[1]{#1_{1}, \dots ,#1_{n}}
\newcommand{\varsk}[2]{#1_{1}, \dots ,#1_{#2}}
\newcommand{\GG}{\mathcal{G}}
\newcommand{\zi}{z_{i}}
\newcommand\floor[1]{\lfloor#1\rfloor}
\newcommand\ceil[1]{\lceil#1\rceil}
\newcommand\independent{\protect\mathpalette{\protect\independenT}{\perp}}
\def\independenT#1#2{\mathrel{\rlap{$#1#2$}\mkern2mu{#1#2}}}
\newcommand{\gap}{

\vspace{3 mm} \noindent}
\newcommand{\smallg}{

\vspace{1 mm} \noindent}
\newcommand{\real}{\mathbb{R}}
\newcommand{\Proof}{\textit{Proof. }}

\newcommand{\Thm}[1]{\textbf{Theorem #1}}
\newcommand{\Prop}[1]{\textbf{Proposition #1   }}
\newcommand{\Exa}[1]{\textbf{Example #1   }}
\newcommand{\Def}[1]{\textbf{Definition #1   }}
\newcommand{\Lem}[1]{\textbf{Lemma #1   }}
\newcommand{\Cor}[1]{\textbf{Corollary #1   }}
\newcommand{\clg}[1]{\lceil{#1} \rceil}
\newcommand{\intinfx}[1]{\int_{-\infty}^{\infty} #1 \text{ }dx}
\newcommand{\intinf}[2]{\int_{-\infty}^{\infty} #1 \text{ }d#2}
\newcommand{\intx}[3]{\int_{#1}^{#2} #3 dx}
\newcommand{\bo}[1]{\textbf{#1}}
\newcommand{\kh}{k_{h}}
\newcommand{\bigb}[2]{\left(\frac{#1}{#2}\right)}
\newcommand{\E}{E}
\def\T{{ \mathrm{\scriptscriptstyle T} }}
\newcommand{\begineq}[1]{\begin{equation*}
    \begin{split}
    #1
    \end{split}
\end{equation*}}
\newcommand{\begineqn}[1]{\begin{equation}
    \begin{split}
    #1
    \end{split}
\end{equation}}
\newcommand{\dist}{\xrightarrow{d}}
\newcommand{\Ysbar}{\bar{Y}_{S}}
\newcommand{\xra}[1]{\overset{#1}{\rightsquigarrow}}
\newcommand{\chris}[1]{{\textcolor{red}{{(Chris:} #1)}}}
\newcommand{\stephen}[1]{{\textcolor{violet}{{(Stephen:} #1)}}}
\newcommand{\andrew}[1]{{\textcolor{blue}{{(Andrew:} #1)}}}
\newcommand{\edwin}[1]{{\textcolor{orange}{{(Edwin:} #1)}}}
\newcommand{\myrule}{\vrule width 3pt}

\renewcommand\theadalign{ptm}
\renewcommand\theadfont{\bfseries}
\renewcommand\theadgape{\Gape[4pt]}
\renewcommand\cellgape{\Gape[4pt]}

\begin{abstract} 
Causal inference from observational data can be viewed as a missing data problem arising from a hypothetical population-scale randomized trial matched to the observational study. This links a target trial protocol with a corresponding generative predictive model for inference, providing a complete framework for transparent communication of causal assumptions and statistical uncertainty  on treatment effects, without the need for counterfactuals. 
The intuitive foundation for the work is that a whole population randomized trial would provide answers to any observable causal question with certainty. Thus, our fundamental problem of causal inference is the missingness of the hypothetical target trial data, which we solve through repeated imputation from a generative predictive model conditioned on the observational data. Causal assumptions map to intuitive conditions on the transportability of predictive models across populations and conditions.  We demonstrate our approach on a real data application to studying the effects of maternal smoking on birthweights using extensions of Bayesian additive regression trees and inverse probability weighting.
\end{abstract}
{\let\thefootnote\relax\footnote{{$*$Electronic address: \texttt{andrew.yiu@stats.ox.ac.uk}}}}
{\let\thefootnote\relax\footnote{{$^1$Department of Statistics, University of Oxford, UK }}}
{\let\thefootnote\relax\footnote{{$^2$Methods, Innovation \& Outreach, Novo Nordisk, Denmark }}}
{\let\thefootnote\relax\footnote{{$^3$Department of Statistics and Data Sciences, University of Texas at Austin, USA}}}
{\let\thefootnote\relax\footnote{{$^4$The Alan Turing Institute, UK}}}


\section{Introduction}

We consider the problem of causal inference using observational data where a treatment variable (intervention) of interest, $T$, is observed across $n$ independent units with associated outcomes, $Y$, and pre-treatment covariates, $X$. We wish to infer statistical uncertainty in the causal treatment effect of $T$ on the outcome $Y$, using a dataset, $Z_{i} = (Y_{i}, T_{i}, X_{i})$ ($i=1,\ldots,n$) obtained from an observational study, and communicate the necessary assumptions in this learning procedure.

Causal analysis of observational data is delicate and open to bias and misinterpretation if not carefully handled. To address this, \cite{Hernan16} considered the hypothetical design of a randomized target trial, matched to the observational study, as a key stage in the causal analysis workflow. Specifying the target trial protocol can highlight issues of potential bias in the observational study and encourage rigour in defining the causal estimand. We take this concept a step further and consider a generative predictive model for the target trial participants that can be used to quantify statistical uncertainty in the read out of the trial if recruiting the  remaining (not yet observed) population. To state another way, we assume that obtaining trial data,  $Z_{n+1:\infty}$, for the remaining population recruited to the randomized target trial would be sufficient to answer the causal question of interest. 
Causal inference can then be viewed as a predictive missing data problem by imputing the experimental data $Z_{n+1:\infty}$ in the remaining population under randomized treatment allocation, conditioned on the observational data $z_{1:n}$. This allows one to estimate causal effects for any observable quantity without the need to introduce counterfactuals or potential outcomes. In practice we assume a large finite target trial such that the resulting treatment effects can be estimated precisely.

We regard the missing data as the $Z_{n+1:\infty}$ under randomization of the $T_{n+1:\infty}$ while assuming the population characteristics do not change from the observational to the experimental regimes, $P(X \mid \mathcal{O})=P(X \mid \mathcal{E})$, whereas standard use of counterfactuals assume the missing data is the potential outcomes that were unobserved within the $Z_{1:n}$. 


The framing of causal analysis as a population missing data problem builds on the work of \cite{Fong21} who developed the idea for non-causal statistical inference, and \cite{Rubin78} who considered a generative model for potential outcomes under a dependent treatment assignment mechanism. A key difference between our proposal and the potential outcome framework in \cite{Rubin78} is that we condition on the observed data without any obligation to model what might have occurred for the observed units if things had been different. Modeling potential outcomes that weren't realized creates the need for counterfactuals, as the analysis is wrapped up in the $n$ observations.  We focus attention on predicting outcomes in new units under randomized interventions, conditional on the observed information from past units. This removes a lot of the existing machinery for causal inference, and is applicable in the situation where observing $Y_{n+1:\infty}$ under randomized interventions would answer the question of interest. 

\subsection{Related work}

Counterfactual quantities occur in most of the causal inference literature, such as potential outcomes in the Rubin causal model (RCM) \citep{Rubin74, Rubin78, Rubin05} and the Finest Fully Randomized Causally Interpretable Structured Tree Graphs model \citep{Robins86, Richardson13}, and structural equations in the Structural Causal Model of \citet{Pearl09}. These frameworks introduce concepts of parallel worlds where in each world individual units will receive a different treatment allocation. In contrast, we highlight that  in situations where obtaining $Y_{n+1:\infty}$ from a randomized trial would be sufficient to answer the causal question, then  counterfactual concepts are unnecessary.

Our work builds on the target trial emulation framework of \cite{Hernan16} and the decision theoretic approach of \cite{Dawid21}. Target trial emulation \citep{Hernan16} was introduced as a guide to approaching  causal inference problems involving observational data, discussed in Section \ref{sec::tte}. The framework provides a roadmap guiding the statistician through various steps in framing a causal analysis towards the identification (design specification) of a target trial matched to the observational study. Following trial identification, inference proceeds using conventional methods such as the RCM to consider counterfactual outcomes in the observational units. We show that this last step is unnecessary, as uncertainty quantification on the causal effect can be provided within the target trial framework using conditional predictive models to simulate outcomes for the remaining, not yet observed, population as if  recruited to the target trial. In an important paper, \cite{Saarela:2020} show using structural equations that an  assumption of exchangeability leads to an equivalence between identifiable causal effects in the RCM and estimates arising from a population under randomized interventions. We relax assumptions of exchangeability, using martingales to construct predictive models for outcomes under a target trial emulation for a future population that support new methods for causal inference while easily allowing for a range of complexities such as  non-compliance (as discussed in Section \ref{sec::extensions}). 


The rest of the paper is as follows. In Section \ref{sec::caus_fram}, we review target trial emulation and develop our formal causal framework. The differences and similarities between our set-up, potential outcomes and structural equation models are highlighted. The predictive viewpoint using martingales leads to new algorithms for causal inference as shown in Section \ref{sec::pred_causal}. We apply these methods in Section \ref{sec::appl} to study the effect of maternal smoking cessation on birthweight.


 
 
\section{Causal framework} \label{sec::caus_fram}

\subsection{Target trial emulation} \label{sec::tte}

Among the controversy and lively debate that surrounds causality, there is perhaps just a single proposition that enjoys universal consensus: causal conclusions can be drawn from randomized experiments, following \cite{Neyman23}. However, the ability to carry out a randomized controlled trial is often hindered by cost, ethics, practicality, and time constraints. Moreover, even if experimental data are available, there are certain types of causal questions that cannot be precisely answered due to insufficient sample size. 

These limitations of randomized experiments motivate causal analyses of observational data. While the validity of causal inference from observational studies is still hotly contested \citep{Hernan18}, it is becoming increasingly accepted that evidence from outside of randomized controlled trials can be valuable for supporting decision-making \citep{Concato20}.

The most common concern regarding observational studies is the lack of treatment randomization, leading to potential biases due to inadequate adjustment for confounding. However, biases can also arise from inappropriate use of the available data, particularly if the analyst fails to precisely articulate their causal query. Target trial emulation \citep{Hernan16, Hernan20} is a systematic framework for eliminating these types of biases. The user is obliged to frame their causal questions by specifying the protocol of an explicit pragmatic trial---matched to the observational study---that they would have liked to carry out. The analysis of the observational data must then emulate the protocol as closely as possible. 

An example of the specification of the target trial protocol components from \citet{Hernan21} can be found in Table \ref{tab::AIDSprotocol} for a study investigating strategies for initiating antiretroviral therapy to treat HIV. The causal question concerns the difference between initiating antiretroviral therapy as soon as possible after the diagnosis of HIV and initiating only once the patient's immunosuppression (indicated by their CD4 cell count) reaches a particular severe threshold.

A fundamental component of the target trial is the specification of the start of follow-up (also known as \textit{time zero}). Misalignment between time zero and the eligibility criteria can create follow-up periods within which the outcome of interest could not have occurred; this is called \textit{immortal time bias}. It is also crucial to align time zero with the time of treatment assignment. Otherwise, the analysis can be afflicted with selection bias, even if there is adequate adjustment for confounding at baseline. For example, if follow-up begins after treatment assignment and the treatment has a positive effect on survival, it is likely that the proportion of susceptible individuals that have died by time zero is higher in the control group. Thus, there would be imbalance between the two groups despite achieving baseline comparability from randomization.

A successful area of application for target trial emulation is the analysis of observational datasets for severe outcomes and underrepresented subpopulations \citep{Petito20, Yland22,Chiu22}. The analysis proceeds in a two-stage process: the findings from the emulation are shown to match previous randomized trials, and then the results are extended to questions that could not be answered by the randomized trials due to insufficient sample size. Target trial emulation has also been effective at providing evidence for urgent queries before randomized trials can be completed, such as investigating the efficacy of COVID-19 vaccine boosters and comparing the efficacy of different vaccines \citep{Barda21, Dickerman22}, as well as  potential drug repurposing \citep{Laifenfeld2021}.

The target trial framework provides a language to connect observational and experimental analyses. However, the current guidance on the statistical inference \citep{Hernan16, Hernan20} following target trial specification is to revert to general causal approaches such as the g-methods \citep{Robins86, RobinsHernan08} and matching \citep{Rubin73, Rosenbaum83}. These approaches are built on a counterfactual framework, requiring a separate language and set of assumptions (e.g. ``consistency''). 

Our goal is to develop a self-contained statistical framework extending target trial emulation that seamlessly links the protocol components and the statistical inference through the use of predictive generative models (see Table \ref{tab::AIDSprotocol}). Besides the obvious advantages for communication and discussion, we avoid the additional machinery of counterfactuals, which demand unnecessarily strong assumptions for the analysis of a randomized experiment. Furthermore, the use of predictive models for inference enables the user to conveniently combine and compare the information from multiple studies.

\begin{table}[h!]
\caption{An example of the protocol components of a target trial emulation (taken from \citet{Hernan21}) for comparing immediate and deferred initiation of antiretroviral therapy to treat HIV. The components are matched with the notation of our proposed framework. 
\label{tab::AIDSprotocol}}
\begin{center}
\resizebox{1\textwidth}{!}{\begin{tabular}{|l |p{0.3\textwidth}p{0.2\textwidth}|p{0.3\textwidth}c|} \hline
\multirow{2}{4em}{Protocol component} & \multicolumn{2}{c||}{Target trial protocol \quad $(F_{T} = \mathcal{E}$)} & \multicolumn{2}{c|}{Emulation in  observational study \quad $(F_{T} = \mathcal{O}$)}\\ \cline{2-5}
 & \multicolumn{1}{c|}{Description} & \multicolumn{1}{c||}{Generative model} & \multicolumn{1}{c|}{Description} & \multicolumn{1}{c|}{Generative model} \\ \hline
Eligibility criteria  & \multicolumn{1}{p{0.3\textwidth}|}{HIV-positive persons $\geq 18$ years of age with no prior use of antiretroviral therapy
and no history of AIDS.} & \multicolumn{1}{p{0.3\textwidth}||}{$X \in {\cal{X}}$: support of pre-treatment covariates}   & \multicolumn{1}{p{0.3\textwidth}|}{Same as for target protocol.

\textit{Required data for each person: age, history
of therapy use, history of AIDS
diagnosis}} & $X \in {\cal{X}}$ \\ \hline
Treatment strategies & \multicolumn{1}{p{0.3\textwidth}|}{Initiation of antiretroviral therapy:\begin{enumerate}[leftmargin=*,topsep=0pt]
\item Immediately
\item When CD4 cell count drops below
500 cells per cubic millimeter.\end{enumerate}} & \multicolumn{1}{c||}{$T$ } & \multicolumn{1}{p{0.3\textwidth}|}{Same as for target protocol. 

\textit{Required data: date of therapy initiation,
clinical measurements of CD4 cell
count.}} & $T$ \\ \hline
Treatment assignment & \multicolumn{1}{p{0.3\textwidth}|}{Eligible persons will be randomly assigned to one strategy and will be aware of which strategy they were assigned to.} & \multicolumn{1}{c||}{$P(T \mid F_{T} = \mathcal{E})$ } & \multicolumn{1}{p{0.3\textwidth}|}{Eligible persons will be assigned to the
strategies with which their data were
compatible at the time of eligibility.} & $P(T \mid X, F_{T} = \mathcal{O})$\\ \hline
Outcome & \multicolumn{1}{p{0.3\textwidth}|}{Death} & \multicolumn{1}{c||}{$Y$} & \multicolumn{1}{p{0.3\textwidth}|}{Same as for target protocol. 

\textit{Required data: date of death during
the study.}} & $Y$\\ \hline
Follow-up & \multicolumn{1}{p{0.3\textwidth}|}{From treatment assignment until death, loss to follow-up, or administrative end of follow-up, whichever occurs first. The analysis must be adjusted for both pre-baseline and post-baseline prognostic factors that are associated with loss to follow-up.} & \multicolumn{1}{p{0.3\textwidth}||}{Ensure that $X$ contains the pre-baseline factors associated with loss to follow-up.} & \multicolumn{1}{p{0.3\textwidth}|}{Same as for target protocol. 

\textit{Required data: date of loss to follow-up.}} & \multicolumn{1}{p{0.3\textwidth}|}{Ensure that $X$ contains the pre-baseline factors associated with loss to follow-up.} \\ \hline
Causal estimand & \multicolumn{1}{p{0.3\textwidth}|}{\begin{itemize}[leftmargin=*,topsep=0pt]
    \item Average intention-to-treat effect (effect of being assigned to treatment)
 \item Average per-protocol effect (effect of receiving treatment as indicated in the
protocol)
\end{itemize}} & \multicolumn{1}{p{0.3\textwidth}||}{For the per-protocol effect: contrasts between functionals of $P(Y \mid T = t, F_{T} = \mathcal{E})$ for different values of $t$.}  & \multicolumn{1}{p{0.3\textwidth}|}{Observational analogue of the per-protocol effect.} & Same as for target protocol.\\ \hline
\end{tabular}}
\end{center}
\end{table}

\subsection{Set-up} \label{sec::setup}

For the remainder of this section, we will develop our formal causal framework, supplemented with comparisons to potential outcome and structural equation models. To aid communication, we will focus on the simplest set-up with a single time-fixed binary treatment and full compliance. Extensions to more complex settings will be discussed in \S\ref{sec::extensions}. Readers less interested in the technical foundations, who wish to prioritize inference and implementation,  might like to skip ahead to \S\ref{sec::pred_causal} after familiarizing themselves with the notation.

Suppose that we have observed the data $Z_{i} = (Y_{i}, T_{i}, X_{i})$ ($i=1,\ldots,n$) from an observational study.
We work in the setting of ``no unmeasured confounders''; that is, the set of measured covariates $X$ are deemed sufficient to adjust for all confounding between the observational treatment assignment and the outcome. This is commonly represented by the causal directed acyclic graph in Figure \ref{fig::ate_dag}. For expositional clarity and avoidance of measure-theoretic technicalities, we will assume that the covariates are discrete; the ideas we present can be easily extended to general covariates in the usual way.

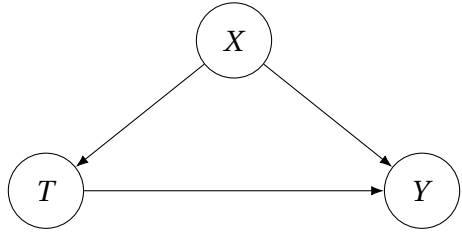
\begin{figure}[]
\begin{center}
\begin{tikzpicture}
\node[state] (x_u) at (0,3) {$X$};
\node[state] (t_u) at (-2.5, 1) {$T$};
\node[state] (y_u) at (2.5, 1) {$Y$};

\path (x_u) edge (t_u);
\path (x_u) edge (y_u);
\path (t_u) edge (y_u);

\end{tikzpicture}
\end{center}
\caption{Standard causal directed acyclic graph for a single time-fixed treatment with no unmeasured confounders. \label{fig::ate_dag}}
\end{figure}

In developing a statistical framework for target trial emulation, we understand that there are two probability distributions of interest: one for the observational regime, and one for the experimental regime. The aim of this subsection is to specify a set of extended conditional independence assumptions \citep{Constantinou17} that will allow us to identify causal estimands in the target trial using just observational data. 

We introduce a regime indicator $F_{T}$, which takes the values $\mathcal{O}$ and $\mathcal{E}$ for the observational regime and the experimental (target trial) regime respectively. This is inspired by the decision-theoretic approach of \citet{Dawid02}, in which the regime indicator takes the values $F_{T} = \emptyset$ and $F_{T} = t$ for each treatment value $t$. The ``idle'' regime $F_{T} = \emptyset$ in \citet{Dawid02} is equivalent to our observational regime $F_{T} = \mathcal{O}$. But the interventional regime $F_{T} = t$ in which the unit is forced to take the treatment $t$ has no direct analogue in our framework. In this article, $F_{T}$ will be non-stochastic; possible extensions to stochastic $F_{T}$---to model non-compliance, for example---will be discussed in \S\ref{sec::extensions}.

Our assumptions will be stated in terms of a Markov kernel $P$---indexed by the two values of $F_{T}$---over a single set of observable variables $(Y, T, X)$. If $Z_{1}, \ldots, Z_{n}$ are assumed to be independent and identically distributed,  $P(\cdot \mid F_{T} = \mathcal{O})$ should be interpreted as the underlying data-generating distribution for a single unit in the observational study; we work within this set-up in \S\ref{sec::clev_cov} and \S\ref{sec::marg_pred}. If the user instead models the data using exchangeability \citep[e.g][]{Dawid21}, then $P(\cdot \mid F_{T} = \mathcal{O})$ corresponds to the ``lurking'' de Finetti parameter for a joint distribution over an infinite exchangeable sequence $(Z_{i})_{i=1}^{\infty}$. In Section \ref{sec::pred_causal}, we will impose the assumptions on a sequence of one-step ahead predictive distributions.

To ensure that the conditional distributions studied below are well-defined, we start by making the following assumption:
\begin{assumption}[Positivity] \label{ass::pos}
The treatment assignments satisfy
\begin{equation*}
    0 < P(T=1 \mid F_{T} = \mathcal{E}) < 1,
\end{equation*}
and with probability 1,
\begin{equation*}
    0 < P(T = 1 \mid X, F_{T} = \mathcal{O}) < 1.
\end{equation*}
\end{assumption}
Since the target trial is hypothetical, and its design is fully under our control, the first part of Assumption \ref{ass::pos}---corresponding to the experimental regime---is technically not an assumption, given that we can simply enforce it to be true.

In the experimental regime $F_{T} = \mathcal{E}$, the fully randomized treatment assignment is independent of everything that occurs prior to the treatment administration. This is expressed in the following assumption.

\begin{assumption}[Randomization] \label{ass::rand}
$T \independent X \mid F_{T} = \mathcal{E}$
\end{assumption}

Similar to the discussion of Assumption \ref{ass::pos}, the hypothetical nature of the target trial implies that Assumption \ref{ass::rand} can be made to hold by design, rather than by assumption.

Assumption \ref{ass::modularity} below links the probability distributions across the two regimes. The key component required to be invariant is the conditional distribution of the outcome given the administered treatment and the covariates; that is, $P(Y \mid T, X, F_{T} = \mathcal{O}) = P(Y \mid T, X, F_{T} = \mathcal{E})$. This links intuitively to our understanding that the presence of confounding is encapsulated by $P(Y \mid T, F_{T} = \mathcal{O}) \neq P(Y \mid T, F_{T} = \mathcal{E})$, or in words, just conditioning on treatment in the observational study will not derive the same results as the target trial. Thus, $X$ is sufficient to adjust for confounding if and only if the equality holds after conditioning further on the covariates.

\begin{assumption}[Modularity] \label{ass::modularity} \leavevmode 
\begin{itemize}
    \item[(i)] $Y \independent F_{T} \mid T,X$
    \item[(ii)] $X \independent F_{T}$
\end{itemize}
\end{assumption}

The assumption of $X \independent F_{T}$ is inappropriate if the target population is believed to differ from the sample population in terms of the distribution of $X$. For average treatment effects, it is still possible to proceed if $P(X \mid F_{T} = \mathcal{E})$---the covariate distribution in the target population---is already known; our framework accommodates this option very intuitively through the use of generative models and also allows for sensitivity analysis under shifting of the population covariate distribution. Otherwise, given the other assumptions, it remains an option to only infer treatment effects that condition on the covariates rather than marginalize them out.

The following result provides the general identification formula for causal estimands in the target trial. It states that $P$ factorises with respect to the augmented directed acyclic graph in Figure \ref{fig::aug_dag}. The square node that contains $F_{T}$ indicates that the variable is non-stochastic, and the dashed edge from $X$ to $T$ is removed upon conditioning on $F_{T} = \mathcal{E}$. The requisite conditional independences follow immediately from Assumptions \ref{ass::rand} and \ref{ass::modularity}.

\begin{figure}[]
\begin{center}
\begin{tikzpicture}

\node[state] (x_u) at (0,3) {$X$};
\node[state] (t_u) at (0, 1) {$T$};
\node[state] (y_u) at (2.5, 1) {$Y$};
\node[square, draw, minimum width = 0.9cm] (f_u) at (-2.5, 1) {$F_{T}$};

\path (x_u) edge (y_u);
\path[dashed] (x_u) edge (t_u);
\path (t_u) edge (y_u);
\path (f_u) edge (t_u);

\end{tikzpicture}
\end{center}
\caption{Augmented causal directed acyclic graph for $(Y,T,X)$ under Assumptions \ref{ass::pos}-\ref{ass::modularity}. \label{fig::aug_dag}}
\end{figure}
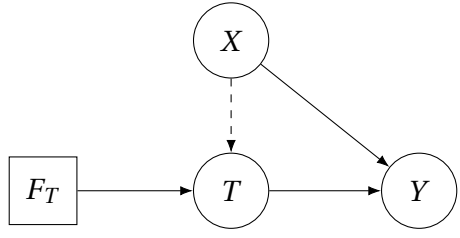

\begin{proposition}[The g-formula density + randomized assignment] \label{prop::gform}
Under Assumptions \ref{ass::pos}-\ref{ass::modularity}, the Markov kernel $P$ factorises with respect to the directed acyclic graph in Figure \ref{fig::aug_dag}. In particular, we have the identification formula
\begin{equation} \label{eqn::gform}
    P(Y,T,X \mid F_{T} = \mathcal{E}) = P(Y \mid T,X,F_{T} = \mathcal{O})P(T \mid F_{T} = \mathcal{E})P(X \mid F_{T} = \mathcal{O}).
\end{equation}
\end{proposition}

This is almost exactly analogous to the ``g-formula'' density \citep{Robins86,Hernan20}, but we have the extra factor $P(T \mid F_{T} = \mathcal{E})$ corresponding to the fully-randomized treatment assignment in the target trial, while the usual g-formula density fixes the administered treatment to a particular value. 

As an immediate corollary of Proposition \ref{prop::gform} (using just the laws of probability), we obtain the identification formula for the conditional outcome distribution given administered treatment in the experimental regime:
\begin{equation} \label{eqn::ate_iden_den}
    P(Y \mid T=t, F_{T} = \mathcal{E}) = \sum_{x} P(Y \mid T = t, X=x, F_{T} = \mathcal{O})P(X=x \mid F_{T} = \mathcal{O})
\end{equation}
for each $t = 0,1$.

\begin{example}
    The average treatment effect is
    \begin{equation*}
        E[Y \mid T=1, F_{T} = \mathcal{E}] - E[Y \mid T=0, F_{T} = \mathcal{E}],
    \end{equation*}
    which is identified as
    \begin{equation} \label{eqn::ate_iden}
        \sum_{x} \left(E[Y \mid T = 1, X=x, F_{T} = \mathcal{O}]-E[Y \mid T = 0, X=x, F_{T} = \mathcal{O}]\right)P(X=x \mid F_{T} = \mathcal{O})
    \end{equation}
    under Assumptions \ref{ass::pos}-\ref{ass::modularity}.
    
    We can also identify the average treatment effect with inverse probability weighting:
    \begin{equation*}
        E[Y \mid T=1, F_{T} = \mathcal{E}] = E\left[\frac{TY}{P(T=1 \mid X, F_{T} = \mathcal{O})} \,\middle\vert\, F_{T} = \mathcal{O}\right],
    \end{equation*}
    with a similar expression for $T=0$. To see this, we take the conditional expectation of the numerator on the right-hand side:
    \begin{equation*}
        E[TY \mid X, F_{T} = \mathcal{O}] = E[Y \mid T=1, X, F_{T} = \mathcal{O}]P(T=1 \mid X, F_{T} = \mathcal{O}).
    \end{equation*}
    After cancelling $P(T=1 \mid X, F_{T} = \mathcal{O})$ from top and bottom, we are left with the first term in (\ref{eqn::ate_iden}).
\end{example}

\begin{example}
    For a binary outcome variable $Y$, the causal risk ratio is identified as
    \begin{equation*}
        \frac{P(Y=1 \mid T=1, F_{T} = \mathcal{E})}{P(Y=1 \mid T=0, F_{T} = \mathcal{E})} = \frac{\sum_{x} P(Y=1 \mid T = 1, X=x, F_{T} = \mathcal{O})P(X=x \mid F_{T} = \mathcal{O})}{\sum_{x} P(Y=1 \mid T = 0, X=x, F_{T} = \mathcal{O})P(X=x \mid F_{T} = \mathcal{O})}.
    \end{equation*}
\end{example}

Assumptions \ref{ass::pos}-\ref{ass::modularity} are insufficient to infer causal estimands that involve the ``natural'' treatment assignment (also called the ``intention to treat'' variable by \citet{Dawid21}), which determines the received treatment in the absence of intervention. This includes the ``average treatment effect on the treated'' in Example \ref{exa::att} below. In order to proceed, the user must additionally assume that it is possible to observe the natural treatment assignment in the target trial. 

Following \citet{Robins07} and \citet{Dawid21}, we split the treatment variable in two: $T^{\mathcal{O}}$ denotes the natural treatment assignment, while $T$ denotes the treatment that is actually administered to the unit. By definition, these two variables coincide in the observational regime, but they may differ in the target trial. To be more explicit, we can write
\begin{equation*}
T = \begin{cases}
        T^{\mathcal{O}}, & \text{if } F_{T} = \mathcal{O}\\
        T^{\mathcal{E}}, & \text{if } F_{T} = \mathcal{E}
        \end{cases}
\end{equation*}
where $T^{\mathcal{E}}$ is the fully randomized treatment assignment in the target trial. The set of observable variables is now $(Y, T, T^{\mathcal{O}}, X)$, and the new causal graph (before augmentation) is shown in Figure \ref{fig::ate_dag_itt}. 

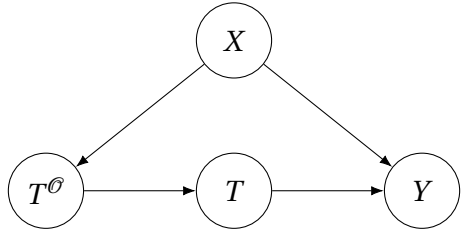
\begin{figure}[]
\begin{center}
\begin{tikzpicture}
\node[state] (x_u) at (0,3) {$X$};
\node[state] (to_u) at (-2.5, 1) {$T^{\mathcal{O}}$};
\node[state] (t_u) at (0, 1) {$T$};
\node[state] (y_u) at (2.5, 1) {$Y$};

\path (x_u) edge (to_u);
\path (to_u) edge (t_u);
\path (x_u) edge (y_u);
\path (t_u) edge (y_u);

\end{tikzpicture}
\end{center}
\caption{Standard causal directed acyclic graph including the natural treatment assignment.\label{fig::ate_dag_itt}}
\end{figure}

Since $T^{\mathcal{O}}$ is assumed to be realised prior to the treatment administration in the target trial, the following holds by design:
\begin{assumption}[Extended randomization] \label{ass::ext_rand}
$T \independent (T^{\mathcal{O}}, X) \mid F_{T} = \mathcal{E}$
\end{assumption}

Next we have an assumption that is analogous to (weak) ``conditional ignorability'' (also known as: ``conditional exchangeability''), which is ubiquitous in potential outcome models \citep[e.g.][]{Hernan20}. In words, it states that the outcome is independent of the natural treatment assignment given the administered treatment and the covariates in the experimental regime. 

\begin{assumption}[Conditional ignorability] \label{ass::cond_ign}
$Y \independent T^{\mathcal{O}} \mid T,X, F_{T}=\mathcal{E}$
\end{assumption}

The final assumption extends the modularity in Assumption \ref{ass::modularity} to $P(T^{\mathcal{O}} \mid X, F_{T})$; we assume that the conditional distribution of the natural treatment assignment is unchanged when we move from the observational regime to the target trial.

\begin{assumption}[Extended modularity] \label{ass::ext_mod}  \leavevmode 
\begin{itemize}
    \item[(i)] $Y \independent F_{T} \mid T,X$
    \item[(ii)] $T^{\mathcal{O}} \independent F_{T} \mid X$
    \item[(iii)] $X \independent F_{T}$
\end{itemize}
\end{assumption}

For readers who are familiar with Pearl's Structural Causal Model \citep{Pearl09}, the functional version of Assumption  \ref{ass::ext_mod} in a nonparametric structural equation model is
\begin{align*}
    X &= U_{X} \\
    T^{\mathcal{O}} &= f_{T^{\mathcal{O}}}(X, U_{T^{\mathcal{O}}}) \\
    Y &= f_{Y}(T,X,U_{Y}),
\end{align*}
where $U_{X}$, $U_{T^{\mathcal{O}}}$ and $U_{Y}$ are jointly independent exogenous variables, and $f_{T^{\mathcal{O}}}$ and $f_{Y}$ are deterministic functions. We point out that the structural model is a strictly stronger assumption since Assumption \ref{ass::ext_mod} is agnostic about the existence of such a functional representation; modularity is only assumed on a distributional level. In particular, the functional form immediately posits the existence of potential outcomes $f(1, X, U_{Y})$ and $f(0,X,U_{Y})$ and a joint probability distribution over both variables. Furthermore, the functional form assumes implicitly that it is possible to intervene on $X$, whereas this is not implied by Assumption \ref{ass::ext_mod}.

We point out that conditional ignorability (Assumption \ref{ass::cond_ign}) was not required for Proposition \ref{prop::gform}. In \S\ref{sec::po}, we will discuss the usual derivation of (\ref{eqn::ate_iden_den}) with potential outcomes using conditional ignorability and consistency.

\begin{figure}[]
\begin{center}
\begin{tikzpicture}
\node[state] (x_u) at (0,3) {$X$};
\node[state] (to_u) at (-2.5, 1) {$T^{\mathcal{O}}$};
\node[state] (t_u) at (0, 1) {$T$};
\node[state] (y_u) at (2.5, 1) {$Y$};
\node[square, draw, minimum width = 0.9cm] (f_u) at (0, -1) {$F_{T}$};

\path (x_u) edge (to_u);
\path (x_u) edge (y_u);
\path[dashed] (to_u) edge (t_u);
\path (t_u) edge (y_u);
\path (f_u) edge (t_u);

\end{tikzpicture}
\end{center}
\caption{Augmented causal directed acyclic graph under Assumptions \ref{ass::pos} and \ref{ass::ext_rand}-\ref{ass::ext_mod}. \label{fig::aug_dag_ext}}
\end{figure}
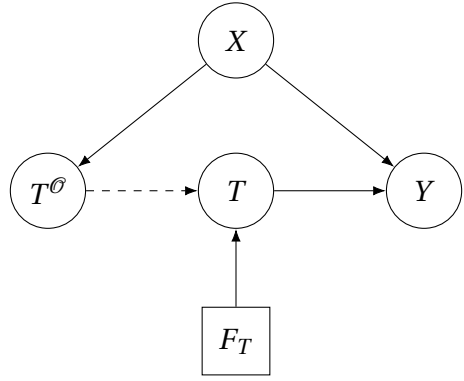

\begin{theorem}[The extended g-formula density + randomized assignment] \label{theo::fact}
Under Assumptions \ref{ass::pos} and \ref{ass::ext_rand}-\ref{ass::ext_mod}, the Markov kernel $P$ factorises with respect to the directed acyclic graph in Figure \ref{fig::aug_dag_ext}. In particular, we have the identification formula
\begin{equation} \label{eqn::ext_gform}
    P(Y,T,T^{\mathcal{O}},X \mid F_{T} = \mathcal{E}) = P(Y \mid T,X,F_{T} = \mathcal{O})P(T \mid F_{T} = \mathcal{E})P(T^{\mathcal{O}} \mid X, F_{T} = \mathcal{O})P(X \mid F_{T} = \mathcal{O}).
\end{equation}
\end{theorem}

Similar to the relationship between Proposition \ref{prop::gform} and the g-formula density, equation (\ref{eqn::ext_gform}) in Theorem \ref{theo::fact} is analogous to the ``extended g-formula'' density \citep{RobinsHernan04, Richardson13} aside from the extra factor $P(T \mid F_{T} = \mathcal{E})$. We can use (\ref{eqn::ext_gform}) to derive the identification formula for the intention-to-treat outcome distribution using only probability calculus. For readers familiar with Single-World Intervention Graphs (SWIG's) \citep{Richardson13}, the separation of $T^{\mathcal{O}}$ and $T$ in Figures \ref{fig::ate_dag_itt} and \ref{fig::aug_dag_ext} may be reminiscent of the ``node-splitting'' transformation under treatment intervention. 

\begin{corollary}[Identification of the intention-to-treat outcome distribution] \label{cor::att_iden}
Under Assumptions \ref{ass::pos} and \ref{ass::ext_rand}-\ref{ass::ext_mod},
\begin{equation*}
    P(Y \mid T=t, T^{\mathcal{O}} = 1,F_{T} = \mathcal{E}) = \sum_{x} P(Y \mid T = t, X=x, F_{T} = \mathcal{O})P(X=x \mid T^{\mathcal{O}} = 1, F_{T} = \mathcal{O})
\end{equation*}
for each $t = 0,1$.
\end{corollary}

\begin{example} \label{exa::att}
The average treatment effect on the treated is
\begin{equation*}
    E[Y \mid T=1, T^{\mathcal{O}} = 1, F_{T} = \mathcal{E}]-E[Y \mid T=0, T^{\mathcal{O}} = 1, F_{T} = \mathcal{E}],
\end{equation*}
which is identified as
    \begin{equation*} 
        \sum_{x} \left(E[Y \mid T = 1, X=x, F_{T} = \mathcal{O}]-E[Y \mid T = 0, X=x, F_{T} = \mathcal{O}]\right)P(X=x \mid T^{\mathcal{O}} = 1, F_{T} = \mathcal{O})
    \end{equation*}
    under Assumptions \ref{ass::pos} and \ref{ass::ext_rand}-\ref{ass::ext_mod}.
\end{example}

\subsection{Comparison with potential outcome models} \label{sec::po}

Within the set-up of \S\ref{sec::setup}, potential outcome models posit the existence of a set of variables $(Y(t), T^{\mathcal{O}}(t), X(t))$ for every treatment value $t$. These variables correspond to the the interventional regimes under which the unit was forced to take a particular treatment. The ``factual'' variables $(Y, T^{\mathcal{O}}, X)$ and all sets of ``counterfactual'' variables above are defined on a common probability space with probability distribution $Q$.

Aside from the SWIG model discussed later, all potential outcome models make the following assumption that links the factual and counterfactual variables on an individual level.

\begin{assumption}[Recursive substitution] \label{ass::recur} \leavevmode 
\begin{itemize}
    \item[(i)] \textit{(Functional consistency)} $T^{\mathcal{O}}=t \implies Y(t) = Y$. 
    \item[(ii)] $T^{\mathcal{O}}(0)=T^{\mathcal{O}}(1)=T^{\mathcal{O}}$
    \item[(iii)] $X(0)=X(1)=X$
    
\end{itemize}
\end{assumption}

If we informally identify $P(T^{\mathcal{O}}, X \mid T=t, F_{T} = \mathcal{E})$ with $Q(T^{\mathcal{O}}(t), X(t))$, we see that the target trial analogues of Assumption \ref{ass::recur}(ii) and \ref{ass::recur}(iii) are Assumptions \ref{ass::ext_mod}(ii) and \ref{ass::ext_mod}(iii) respectively. Usually, Assumption \ref{ass::recur}(i) is simply referred to as ``consistency''. We can describe this as: ``If the individual would take treatment $t$ in the absence of intervention, the observed outcome $Y$ is exactly the same as the outcome that would have been observed if they were instead forced to take treatment $t$.''

\begin{table}[h!]
\caption{Rubin's potential outcomes as missing data. The potential outcome $Y_{i}(1-T^{\mathcal{O}})$ is counterfactual on treatment allocation. \label{tab::rubin}}
\begin{center}
\resizebox{.4\textwidth}{!}{\begin{tabular}{|l |cccc|} \hline
Unit & $Y(0)$ & $Y(1)$ & $T^{\mathcal{O}}$ & $X$ \\ \hline
1  & ? & $Y_{1}(1)$ & 1 & $X_{1}$\\
2  & $Y_{2}(0)$ & ? & 0 & $X_{2}$  \\
$\vdots$ & $\vdots$ & $\vdots$ & $\vdots$ & $\vdots$\\
$n$  & ? & $Y_{n}(1)$ & 1 & $X_{n}$\\ \hline
\end{tabular}}
\end{center}
\end{table}

Consistency is crucial to the ``missing data'' interpretation of the potential outcome framework \citep{Rubin74, Rubin78} illustrated in Table \ref{tab::rubin}. However, consistency is unverifiable from any target trial; it is possible for it to be false and yet unfalsifiable. This is because we can never observe both $Y$ and $Y(t)$ for a single unit. The falsifiable component is captured in the following assumption introduced by \citet{Richardson22} that replaces functional consistency in the SWIG potential outcome model.

\begin{assumption}[Distributional consistency for potential outcomes] \label{ass::dcons_po} For each $t = 0, 1$,
\begin{equation*}
    Q(Y(t)=y, T^{\mathcal{O}}(t) = t, X(t)=x) = Q(Y=y, T^{\mathcal{O}} = t, X=x).
\end{equation*}
\end{assumption}

This weaker version of consistency comes at the cost of losing the missing data interpretation, leaving little motivation for proceeding in a potential outcome framework. We can restate distributional consistency in the target trial framework (see also: Definition 2 in \citet{Dawid21}).

\begin{assumption}[Distributional consistency for target trial emulation] \label{ass::dcons_hre} For each $t = 0, 1$,
\begin{equation*}
    P(X,Y \mid T^{\mathcal{O}} = t, T=t, F_{T} = \mathcal{E}) = P(X,Y \mid T=t, F_{T} = \mathcal{O}).
\end{equation*}
\end{assumption}

This states that if the observational and randomized treatment assignments happen to take the same value in the experimental regime, the distribution of $(X,Y)$ is equal to their distribution under the observational regime given that common treatment value. Assumption \ref{ass::dcons_hre} is implied by Assumptions \ref{ass::pos} and \ref{ass::ext_rand}-\ref{ass::ext_mod}; this follows from Theorem \ref{theo::fact} and observing that $(T, T^{\mathcal{O}})$ d-separates $(X,Y)$ from $F_{T}$ in Figure \ref{fig::aug_dag_ext}. It is also straightforward to see that conditional ignorability (Assumption \ref{ass::cond_ign}) and distributional consistency imply Assumption \ref{ass::ext_mod}(i):
\begin{equation*}
    P(Y \mid T=t,X, F_{T} = \mathcal{E}) = P(Y \mid T^{\mathcal{O}} = t,T=t,X, F_{T} = \mathcal{E}) = P(Y \mid T=t,X, F_{T} = \mathcal{O}),
\end{equation*}
where the first equality is due to conditional ignorability, and the second follows from distributional consistency. Thus, it is possible to replace Assumption \ref{ass::ext_mod}(i) with distributional consistency to obtain Theorem 1, which closely matches the approach of \citet{Dawid21} and the usual potential outcome set-up.

We have chosen to use Assumption \ref{ass::ext_mod}(i) in our set-up for two reasons. First it allows us to clearly delineate between the different settings in Proposition \ref{prop::gform} and Theorem \ref{theo::fact}, depending on whether $T^{\mathcal{O}}$ is observable in the target trial. Moreover, as discussed previously, Assumption \ref{ass::ext_mod}(i) intuitively captures the notion that $X$ is sufficient to adjust for confounding, and it is more interpretable than consistency, which is unfortunately associated with multiple different explanations \citep{Robins95, Cole09, VanderWeele09, Pearl10, Dawid21}.

Along with Assumption \ref{ass::recur}, the potential outcomes framework makes the following two assumptions.

\begin{assumption}[Positivity for potential outcomes] \label{ass::pos_po}
With probability 1, $0 < Q(T^{\mathcal{O}}=1 \mid X) < 1$.
\end{assumption}

\begin{assumption}[Conditional ignorability for potential outcomes] \label{ass::con_ign_po} For each $t$, $Y(t) \independent T^{\mathcal{O}} \mid X$.
\end{assumption}

The extended g-formula density is a simple consequence of Assumptions \ref{ass::recur}, \ref{ass::pos_po} and \ref{ass::con_ign_po}:
\begin{equation*}
    Q(Y(t),T^{\mathcal{O}}(t),X(t)) = Q(Y \mid T^{\mathcal{O}} = t, X)Q(T^{\mathcal{O}} \mid X)Q(X).
\end{equation*}

We argue that the separation of the treatment variable into the natural assignment $T^{\mathcal{O}}$ and the administered treatment $T$ makes Assumption \ref{ass::cond_ign} far easier to understand than Assumption \ref{ass::con_ign_po}, which is often misinterpreted as a conditional independence between the observed outcome and the administered treatment (see, for example, p. 53 of \citet{Morgan07}). 

Finally, we point out that ``strong ignorability''---$(Y(0),Y(1)) \independent T^{\mathcal{O}} \mid X$---has no analogue in our framework. Like the DT framework in \citet{Dawid21}, our framework automatically protects the user from stating ``cross-world'' conditional independence assumptions \citep{Richardson13} that cannot be verified in any target trial.

\begin{table}[h!]
\caption{Comparison of assumptions between the potential outcomes and target trial frameworks.  \label{tab::assum}}
\begin{center}
\resizebox{.8\textwidth}{!}{\renewcommand{\arraystretch}{2}{\begin{tabular}{|l |c|c| p{0.4\textwidth}c|}
\hline & \makecell{Potential outcomes} & Target Trial Predictive  & \multicolumn{1}{c|}{Target Trial Descriptive}\\ \hline
Positivity & \makecell{With probability 1,\\
$0 < Q(T=1 \mid X) < 1.$} & \makecell{With probability 1,\\
$0 < P(T=1 \mid X, F_{T} = \mathcal{O}) < 1.$} & \multicolumn{1}{p{0.4\textwidth}|}{The conditional probability of receiving every value of treatment within levels of the pre-treatment covariates is greater than zero in the observational study.}\\ \hline
Ignorability & \makecell{For each $t=0,1$,\\
$Y(t) \independent T \mid X$.} & \makecell{$Y \independent T^{\mathcal{O}} \mid (T,X, F_{T}=\mathcal{E})
$} & \multicolumn{1}{p{0.4\textwidth}|}{The natural treatment assignment is conditionally independent of the outcome in the target trial given the administered treatment and pre-treatment covariates.}\\\hline
Consistency & \makecell{For each $t=0,1$,\\
$T=t \implies Y(t) = Y$.} & \makecell{For each $t=0,1$,\\
$Y \independent F_{T} \mid T=T^{\mathcal{O}} = t$.}  & \multicolumn{1}{p{0.4\textwidth}|}{If an individual receives treatment $t$, then the  \textit{conditional distribution} of the outcome is unchanged across assignment regimes. This assumption is redundant for the target trial framework, since it is implied by the ignorability condition (see above) and the modularity condition (see below). }\\\hline
Modularity & \makecell{For each $t=0,1$,\\
$Q(Y(t) | X) = Q(Y \mid T=t, X)$,\\ $(T^{\mathcal{O}}(t),X(t))=(T^{\mathcal{O}},X).$} & \makecell{$Y \independent F_{T} \mid T,X$\\
$(T^{\mathcal{O}},X) \independent F_{T}$.} & \multicolumn{1}{p{0.4\textwidth}|}{The predictive distributions characterising the statistical relationships between the variables are \textit{transportable} across regimes. This assumption can be relaxed to allow for changes in the predictive distributions across regimes, while noting such changes are untestable without further information.} \\\hline
\end{tabular}}}
\end{center}
\end{table}

\subsection{Extensions} \label{sec::extensions}

Thus far, we have restricted our attention to the simplest setting of a single binary time-fixed treatment and full compliance. In this subsection, we provide some suggestions on extending the set-up.

For complex longitudinal settings, the most straightforward way to derive the requisite conditional independencies is to take a graphical approach. Starting from an observational graph, the augmentation proceeds in two steps. First, a regime indicator is introduced for each intervenable variable; these indicators are founder nodes in the augmented graph, each with a single edge going into their corresponding variables. Second, all other edges going into the intervenable variables become dashed, signifying that those edges should be removed when the associated regime indicator is switched from $\mathcal{O}$ to $\mathcal{E}$. An example that was studied in \citet{Richardson13} and \citet{Dawid21} is illustrated in Figure \ref{fig::dag_long}; the model contains a latent variable $H$ and two treatments $T_{0}$ and $T_{1}$.

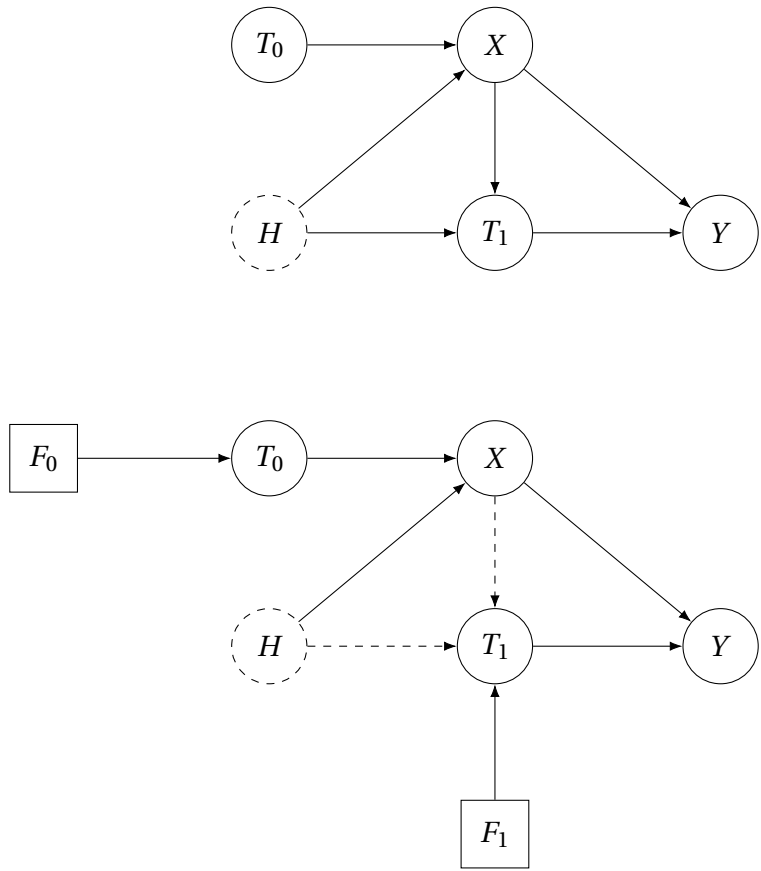
\begin{figure}[]
\begin{center} 
\begin{tikzpicture}
\node[state] (x_u) at (0,3.5) {$X$};
\node[state] (t_0) at (-3,3.5) {$T_{0}$};
\node[state, dashed] (h) at (-3,1) {$H$};
\node[state] (t_1) at (0, 1) {$T_{1}$};
\node[state] (y_u) at (3, 1) {$Y$};

\node[state] (1x_u) at (0,-2) {$X$};
\node[state] (1t_0) at (-3,-2) {$T_{0}$};
\node[state, dashed] (1h) at (-3,-4.5) {$H$};
\node[state] (1t_1) at (0, -4.5) {$T_{1}$};
\node[state] (1y_u) at (3, -4.5) {$Y$};
\node[square, draw, minimum width = 0.9cm] (1f_1) at (0, -7) {$F_{1}$};
\node[square, draw , minimum width = 0.9cm] (1f_0) at (-6, -2) {$F_{0}$};

\path (t_0) edge (x_u);
\path (x_u) edge (t_1);
\path (x_u) edge (y_u);
\path (t_1) edge (y_u);
\path (h) edge (t_1);
\path (h) edge (x_u);

\path (1t_0) edge (1x_u);
\path[dashed] (1x_u) edge (1t_1);
\path (1x_u) edge (1y_u);
\path (1t_1) edge (1y_u);
\path[dashed] (1h) edge (1t_1);
\path (1h) edge (1x_u);
\path (1f_1) edge (1t_1);
\path (1f_0) edge (1t_0);

\end{tikzpicture}
\end{center}
\caption{Observational graph (top) and augmented graph (bottom) for a longitudinal setting.\label{fig::dag_long}}
\end{figure}

For a particular treatment variable, any other variable given its history will be blocked from the treatment regime indicator by construction. Therefore, the necessary modularity conditions ---like Assumptions \ref{ass::modularity} and \ref{ass::ext_mod}---are automatically induced by the addition of the regime indicators. The dashed edges encode the randomization assumptions that stipulate the independence between any treatment variable and its past in the experimental regime. The appropriate extensions of positivity are clear.

Partial compliance in a target trial can be modeled by generalizing the regime indicator to be a stochastic variable. More specifically, the regime $F_{T} = \mathcal{O}$ can represent the situation in which the individual ignores their assigned treatment and chooses based on their free will. In this case, $P$ is now a full joint distribution over all variables, including the regime indicators. The user is additionally required to model the marginal distribution of $F_{T}$, representing the probability of compliance. We leave the specification details to future work. The key point is that extensions are handled naturally within the target trial framework by extending the generative models to capture the additional complexities.

\section{Predictive causal inference} \label{sec::pred_causal}

\subsection{Target trial predictive resampling}

Despite the absence of counterfactuals in our set-up, we concur that causal inference is fundamentally a missing data problem solvable with standard notions of conditional probability. This view is based on the primitive notion that a population-scale target trial would allow us to answer our causal question with practical certainty. Thus, our uncertainty flows entirely from the ``missing'' target trial data, which we label with $k \in \{n+1,\ldots, N\}$. This perspective is illustrated in Table \ref{tab::target_tab}, which should be contrasted with Rubin's potential outcomes table in Table \ref{tab::rubin}.

\begin{table}[h!]
\caption{Target trial emulation as missing data, imputed using a generative model $p(Y,T,T^{\mathcal{O}},X)$. \label{tab::target_tab}}
\begin{center}
\resizebox{.4\textwidth}{!}{\begin{tabular}{|l |cccc|c|} \hline
Unit & $Y$ & $T$ & $T^{\mathcal{O}}$ & $X$ & $F_{T}$\\ \hline
1   & $Y_{1}$ & 1 & 1 & $X_{1}$ & $\mathcal{O}$\\
2  & $Y_{2}$ & 0 & 0 & $X_{2}$ & $\mathcal{O}$ \\
$\vdots$ & $\vdots$ & $\vdots$ & $\vdots$ & $\vdots$ & $\vdots$\\
$n$   & $Y_{n}$ & 1 & 1 & $X_{n}$ & $\mathcal{O}$\\ \hline
$n+1$ & ? & ? & ? & ? & $\mathcal{E}$ \\
$n+2$ & ? & ? & ? & ? & $\mathcal{E}$\\
$\vdots$ & $\vdots$ & $\vdots$ & $\vdots$ & $\vdots$ & $\vdots$\\
\end{tabular}}
\end{center}
\end{table}

For the purpose of exposition, we will assume throughout this section that $T^{\mathcal{O}}$ is observed in the target trial; in this case, the target trial data that are yet to be observed are $Z_{n+1:N}$, where $Z_{k} = (Y_{k}, T_{k}, T^{\mathcal{O}}_{k}, X_{k})$. 
It will be straightforward to deduce the simpler setting where $T^{\mathcal{O}}$ is possibly unobserved. 

Suppose that each $Z_{k}$ takes values in a measurable space $(S, \mathcal{B})$. Our task is to elicit a joint predictive distribution on $Z_{n+1:N}$ given the data $z_{1:n}$ from an observational study, which can be achieved by specifying a sequence of one-step ahead predictive distributions. More specifically, let $P_{n}$ be a probability measure on $\mathcal{B}$, and for each $k \geq n+1$, let $P_{k}$ be a Markov kernel from $(S^{k-n}, \mathcal{B}^{k-n})$ to ($S, \mathcal{B})$, which induces the law
\begin{equation*}
    \mathcal{L}(Z_{n+1:N}) = \bigotimes_{k=n+1}^{N}P_{k-1}.
\end{equation*}
This remains valid in the limit as $N \rightarrow \infty$ 
due to the Ionescu-Tulcea theorem, which states that there exists a unique probability distribution on $Z_{n+1:\infty}$ corresponding to each infinite sequence of predictives $\{P_{k}:k \geq n\}$.

Let $\mathcal{G}_{n}$ be the trivial sigma-field and let $\mathcal{G}_{k} = \sigma(Z_{n+1:k})$ for $k \geq n+1$. We require our sequence of predictives to produce a \textit{conditionally identically distributed} (c.i.d.) sequence $Z_{n+1}, Z_{n+2},\ldots$ \citep{Berti04}:
\begin{align*}
    \text{Pr}(Z_{k+i} \in \cdot \mid \mathcal{G}_{k-1}) = \text{Pr}(Z_{k} \in \cdot \mid \mathcal{G}_{k-1})= P_{k-1}(\cdot) 
\end{align*}
for all $i > 0$ and $k \geq n+1$. In words, all future observations are identically distributed given the past. The c.i.d. property is a weakening of exchangeability \citep{Kallenberg88}.

Consequently, the predictives satisfy the two \textit{predictive coherence} conditions of \citet{Fong21}; namely, there exists a random probability measure $P_{\infty}$ on $(S, \mathcal{B})$ satisfying
\begin{equation*}
    P_{N}(B) \rightarrow P_{\infty}(B) \quad \text{as }N \rightarrow \infty
\end{equation*}
and
\begin{equation*}
    E[P_{\infty}(B)] = P_{n}(B)
\end{equation*}
almost surely for every fixed $B \in \mathcal{B}$. Furthermore, we have
\begin{equation*}
    \mathbb{P}_{N}(B) \rightarrow P_{\infty}(B) \quad \text{as }N \rightarrow \infty
\end{equation*}
almost surely, where
\begin{equation*}
    \mathbb{P}_{N} = \frac{1}{N-n}\sum_{k = n+1}^{N}\delta_{Z_{k}}
\end{equation*}
is the empirical measure \citep{Berti04, Fong21}. 

We also impose some further restrictions on the predictives to capture the structure of the target trial. First, we require each $P_{k}$ to factor according to the graph in Figure $\ref{fig::aug_dag_ext}$ in the experimental regime $F_{T} = \mathcal{E}$; that is
\begin{equation} \label{eqn::pred_fact}
    P_{k}(Y,T,T^{\mathcal{O}},X) = P_{k}(Y \mid T,X)P_{k}(T)P_{k}(T^{\mathcal{O}} \mid X)P_{k}(X)
\end{equation}
Moreover, to reflect the fact that the treatment assignment in the target trial is under full experimental control, we enforce the sequence $T_{n+1}, T_{n+2},\ldots$ to be independent and identically distributed. For concreteness, we fix $T_{k} \sim \text{Ber}(0.5)$ independently of the past, representing the flip of a fair coin to assign units to treatment or control.

By defining our target estimand as a functional $\theta_{\infty} = \theta(P_{\infty}$) of the limiting predictive, the distribution on $P_{\infty}$ induces the \textit{martingale posterior} \citep{Fong21}
\begin{equation*}
    \Pi_{\infty}(\theta_{\infty} \in A) \coloneqq \int 1(\theta(P_{\infty}) \in A)\Pi(dP_{\infty})
\end{equation*}
for every measurable set $A$. For finite $N$, we can replace $P_{\infty}$ with either $P_{N}$ or $\mathbb{P}_{N}$.

Given a sequence of predictives, we can draw from the martingale posterior by repeatedly imputing the missing target trial data through the computational scheme
\begin{align*}
    Z_{n+1} &\sim P_{n}\\
    Z_{n+2}\mid Z_{n+1} &\sim P_{n+1}\\
    &\vdots\\
    Z_{N} \mid Z_{n+1:N-1} &\sim P_{N-1},
\end{align*}
which was introduced and named \textit{predictive resampling} by \citet{Fong21}. 

Subsequently, the imputed $Z_{n+1:N}$ can be analysed in the same manner as a dataset from a randomized trial. For example, one could compute Neyman's difference-in-means estimator of the average treatment effect
\begin{equation} \label{eqn::ateN}
\begin{split}
    \theta^{ATE}_{N} &= E_{\mathbb{P}_{N}}[Y \mid T=1, F_{T} = \mathcal{E}] - E_{\mathbb{P}_{N}}[Y \mid T=0, F_{T} = \mathcal{E}]\\
    &=\frac{\sum_{k=n+1}^{N}Y_{k}1(T_{k} = 1)}{\sum_{k=n+1}^{N}1(T_{k} = 1)}-\frac{\sum_{k=n+1}^{N}Y_{k}1(T_{k} = 0)}{\sum_{k=n+1}^{N}1(T_{k} = 0)},
    \end{split}
\end{equation}
and similarly, the difference-in-means estimator of the average treatment effect on the treated
\begin{equation} \label{eqn::attN}
\begin{split}
    \theta^{ATT}_{N} &= E_{\mathbb{P}_{N}}[Y \mid T=1, T^{\mathcal{O}} = 1, F_{T} = \mathcal{E}] - E_{\mathbb{P}_{N}}[Y \mid T=0, T^{\mathcal{O}} = 1, F_{T} = \mathcal{E}]\\
    &= \frac{\sum_{k=n+1}^{N}Y_{k}1(T_{k} = 1, T^{\mathcal{O}}_{k} = 1)}{\sum_{k=n+1}^{N}1(T_{k} = 1,T^{\mathcal{O}}_{k} = 1)}-\frac{\sum_{k=n+1}^{N}Y_{k}1(T_{k} = 0, T^{\mathcal{O}}_{k} = 1)}{\sum_{k=n+1}^{N}1(T_{k} = 0, T^{\mathcal{O}}_{k} = 1)}.
    \end{split}
\end{equation}

If we are interested in the limit as $N \rightarrow \infty$, the c.i.d. property justifies truncating the predictive resampling at a large value of $N$ and using $P_{N}$ or $\mathbb{P}_{N}$ as a finite sample approximation to $P_{\infty}$. The factorisation in (\ref{eqn::pred_fact}) offers a natural sequential sampling strategy; pseudo-code for our target trial predictive resampling algorithm can be found in Algorithm \ref{alg::tt_predsamp}.

\begin{algorithm}
\caption{Target trial predictive resampling}\label{alg::tt_predsamp}
Specify $P_{n}$ using the observed data $z_{1:n}$\;
$N > n$ is a large integer\;
 \For{$j \gets 1$ to $B$}{
  \For{$k \gets n+1$ to $N$}{
   Sample $X_{k}$ from $P_{k-1}(X)$ \textit{(Predict the pre-treatment covariates for the $k$-th unit)}\;
   Sample $T^{\mathcal{O}}_{k}$ from $P_{k-1}(T^{\mathcal{O}} \mid X_{k})$ \textit{(Predict the natural treatment assignment that $X_{k}$ would take without an intervention)}\;
   Sample $T_{k}$ from $\text{Ber}(0.5)$ \textit{(Assign a randomized treatment)}\;
   Sample $Y_{k}$ from $P_{k-1}(Y \mid T_{k}, X_{k})$ \textit{(Predict the outcome given the administered treatment and pre-treatment covariates)}\;
   Update the predictive to $P_{k} \mid X_k, T_k, Y_k$\;
   }
   Compute $\mathbb{P}_{N}$ from $\{Z_{n+1:N}\}$\;
   Evaluate $\theta_{N}^{(j)} = \theta(P_{N})$ or $\theta_{N}^{(j)} = \theta(\mathbb{P}_{N})$\;
 }
 Return $\{\theta_{N}^{(1)}, \ldots , \theta_{N}^{(B)}\}$.
\end{algorithm}

\subsection{Predictive specification} \label{sec::pred_spec}

In the previous subsection, we introduced a general scheme to derive martingale posteriors for our target causal estimands. We now provide guidance on eliciting a sequence of predictives that generates c.i.d. data. A conspicuous problem is the absence of target trial data with which to build our first-step predictive $P_{n}$. To resolve this, we instead elicit a first-step predictive for the observational regime and transport it into the experimental regime using the set-up in \S\ref{sec::setup}.

More explicitly, let $P_{n}(\cdot \mid F_{T} = \cdot)$ be a Markov kernel indexed by $F_{T} \in \{\mathcal{O}, \mathcal{E}\}$. By enforcing $P_{n}$ to satisfy Assumptions \ref{ass::pos} and \ref{ass::ext_rand}-\ref{ass::ext_mod}, Theorem \ref{theo::fact} implies that $P_{n}$ factorizes with respect to Figure \ref{fig::aug_dag_ext}, and the first-step predictive for the experimental regime is given by the extended g-formula density
\begin{equation*}
    P_{n}(Y,T,T^{\mathcal{O}},X \mid F_{T} = \mathcal{E}) = \underbrace{P_{n}(Y \mid T,X,F_{T} = \mathcal{O})}_{\circled{A}}\underbrace{P_{n}(T \mid F_{T} = \mathcal{E})}_{\circled{B}}\underbrace{P_{n}(T^{\mathcal{O}} \mid X, F_{T} = \mathcal{O})}_{\circled{C}}\underbrace{P_{n}(X \mid F_{T} = \mathcal{O})}_{\circled{D}}.
\end{equation*}
Thus, it suffices to elicit $P_{n}(\cdot \mid F_{T} = \mathcal{O})$ from the observational data. For clarity, we recall that the factors on the right-hand side are
\begin{itemize}
    \item[\circled{A}:] the conditional predictive for the outcome given the administered treatment and the pre-treatment covariates (analogous to the ``outcome regression'' model).
    \item[\circled{B}:] the fully randomized treatment assignment in the target trial, which we have taken to be $\text{Ber}(0.5)$ for a binary treatment.
    \item[\circled{C}:] the conditional predictive for the natural treatment assignment given the pre-treatment covariates (analogous to the ``propensity score'' model).
    \item[\circled{D}:] the marginal predictive for the pre-treatment covariates.
\end{itemize}
In \S\ref{sec::marg_pred}, we propose a different approach in which the experimental regime predictive is specified directly.

The next step is to specify a sequence of updates that satisfies the c.i.d. property. Following \citet{Fong21}, we use the following equivalent formulation of c.i.d. sequences in terms of a ``martingale condition'':
\begin{equation*}
    E[P_{k}(Y,T,T^{\mathcal{O}},X) \mid \mathcal{G}_{k-1}] = P_{k-1}(Y,T,T^{\mathcal{O}},X)
\end{equation*}
almost surely for all $k \geq n+1$.

Since the factors on the right of (\ref{eqn::pred_fact}) correspond to separate causal mechanisms, it is natural to specify the update for each factor in isolation such that the c.i.d. property is preserved if the updates for the other factors are modified. One could interpret this as a predictive analogue of the modularity assumptions in \S\ref{sec::setup}. In the following, we shall give sufficient conditions for this criterion to be satisfied.

\begin{assumption} \label{ass::comp_pred}For each $k\geq n+1$,
\begin{enumerate} 
    \item[(i)] $P_{k}(X)$ is $\sigma(X_{n+1:k})$-measurable, and $E[P_{k}(X) \mid \mathcal{G}_{k-1}] = P_{k-1}(X)$ almost surely.
    \item[(ii)] $P_{k}(T^{\mathcal{O}} \mid X)$ is $\sigma(T^{\mathcal{O}}_{n+1:k},X_{n+1:k})$-measurable, and $E[P_{k}(T^{\mathcal{O}} \mid X) \mid X_{k}, \mathcal{G}_{k-1}] = P_{k-1}(T^{\mathcal{O}} \mid X)$ almost surely.
    \item[(iii)] $P_{k}(Y \mid T, X)$ is $\sigma(Y_{n+1:k}, T_{n+1:k},X_{n+1:k})$-measurable, and $E[P_{k}(Y \mid T, X) \mid T_{k}, X_{k}, \mathcal{G}_{k-1}] = P_{k-1}(Y \mid T,X)$ almost surely.
\end{enumerate}
\end{assumption}

\begin{theorem} \label{theo::pred_comp}
Suppose that the sequence of predictives $\{P_{k}: k \geq n\}$ satisfies Assumption \ref{ass::comp_pred}. Then $Z_{n+1}, Z_{n+2},\ldots$ is a c.i.d. sequence.
\end{theorem}

The conditions in Assumption \ref{ass::comp_pred} are very natural. First, it is required that each sequence of conditional predictives only depends on the variables contained in the factor, e.g. $P_{k}(T^{\mathcal{O}} \mid X)$ depends on the data for $(T^{\mathcal{O}}, X)$ but not $Y$. Next, the update for each factor satisfies the corresponding conditional martingale property.

A convenient choice for $\{P_{k}(X):\,k \geq n\}$ is the \textit{Bayesian bootstrap} \citep{Rubin81}, where each one-step ahead predictive is simply the empirical measure given the past:
\begin{equation*}
    P_{k}(X) = \frac{1}{k}\sum_{i=1}^{k}\delta_{X_{i}}(X).
\end{equation*}
In this case, the martingale posterior can be computed directly via
\begin{equation*}
    P_{\infty}(X) = \sum_{i=1}^{n}w_{i}\delta_{X_{i}}(X),
\end{equation*}
where $(w_{1}, \ldots, w_{n}) \sim \text{Dirichlet}(1,\ldots,1)$ is a vector of uniform Dirichlet weights. Assumptions \ref{ass::comp_pred}(ii)  and \ref{ass::comp_pred}(iii) are satisfied by the conditional copula updates in \S4.4.2 and \S4.4.3 of \citet{Fong21}. 

We show below that Assumption \ref{ass::comp_pred} is also satisfied by a Bayesian model with independent priors on each factor.

\begin{example}[Bayesian inference with independent priors] \label{exa::bayes}
We model $Z_{n+1:\infty}$ as exchangeable, enforcing the priors on $P(Y \mid T,X), P(T^{\mathcal{O}} \mid X), P(X)$ to be jointly independent. Following the discussion at the beginning of the subsection, the obvious way to specify the priors for the experimental regime is to derive independent posteriors using the observational data. 

For each $m \geq n+1$, the likelihood function for $Z_{n+1:m}$ is
\begin{equation*}
    \mathcal{L}(\alpha, \beta, \gamma) = \prod_{k=n+1}^{m} P(Y_{k} \mid T_{k}, X_{k}, \alpha)P(T^{\mathcal{O}}_{k} \mid X_{k}, \beta)P(X_{k}, \gamma),
\end{equation*}
where we have omitted the known constant factors $P(T_{k})$. Now specify jointly independent priors
\begin{equation*}
    \Pi(\alpha, \beta,\gamma) = \Pi(\alpha)\Pi(\beta)\Pi(\gamma),
\end{equation*}
which induces jointly independent posteriors
\begin{equation*}
    \Pi(\alpha, \beta,\gamma \mid \mathcal{G}_{m}) = \Pi(\alpha\mid \mathcal{G}_{m})\Pi(\beta\mid \mathcal{G}_{m})\Pi(\gamma \mid \mathcal{G}_{m}).
\end{equation*}

It follows that
\begin{equation*}
    P_{m}(Y \mid T,X) = \int P(Y \mid T,X,\alpha)d\Pi(\alpha \mid \mathcal{G}_{m}),
\end{equation*}
which is clearly $\sigma(Y_{n+1:m}, T_{n+1:m},X_{n+1:m})$-measurable. Furthermore,
\begin{align*}
    E[P_{m}(Y \mid T,X) \mid T_{m}, X_{m}, \mathcal{G}_{m-1}] &= E\left[\int P(Y \mid T,X,\alpha)d\Pi(\alpha \mid \mathcal{G}_{m}) \mid T_{m}, X_{m}, \mathcal{G}_{m-1}\right]\\
    &= \int P(Y \mid T,X,\alpha) E\left[\frac{P(Y_{m} \mid T_{m},X_{m},\alpha)}{P_{m-1}(Y_{m} \mid T_{m}, X_{m})} \mid T_{m}, X_{m},\mathcal{G}_{m-1}\right]d\Pi(\alpha \mid \mathcal{G}_{m-1}).
\end{align*}
The second equality follows by Fubini's theorem, and since $Y_{m} \mid T_{m}, X_{m},\mathcal{G}_{m-1} \sim P_{m-1}(Y_{m} \mid T_{m}, X_{m})$, the conditional expectation of the likelihood ratio is equal to one. Therefore, we have equality with $P_{m-1}(Y \mid T,X)$, and the conditional martingale condition is satisfied. The other components of Assumption \ref{ass::comp_pred} are verified similarly.
\end{example}

\subsection{BART with the clever covariate} \label{sec::clev_cov}

Bayesian additive regression trees (BART)---introduced by \citet{Chipman10}---is a nonparametric regression model that often outperforms state-of-the-art machine learning methods like Random Forests \citep{Breiman01} and gradient boosting \citep{Friedman01}. The performance of BART has been particularly impressive when applied to causal inference \citep{Hill11, Dorie19}, where it is used to model the conditional distribution of the outcome given the administered treatment and the pre-treatment covariates.

It has been demonstrated, however, that the performance of nonparametric methods for causal inference can be substantially improved by tailoring the model to the specific target estimand \citep{Dorie19}. These adjustments generally involve the propensity score \citep{Ray19, Ray20, Hahn20} and do not fit within a traditional Bayesian set-up. We will introduce a modification to BART for estimating the average treatment effect that is well-motivated under our predictive framework, illustrating the added flexibility on offer.

The BART model for continuous outcomes posits a sum-of-trees regression function with mean-zero additive errors:
\begin{equation} \label{eqn::cont_out}
    Y_{i} = \sum_{k=1}^{K} g_{k}(W_{i}) + \varepsilon_{i}, 
\end{equation}
where $\varepsilon_{i} \sim \mathcal{N}(0,\sigma^{2})$ and each summand $g_{k}(w)$ is a piecewise constant function defined by a regression tree, which partitions the covariate space through a set of internal decision nodes. The terminal nodes $(\mathcal{A}_{k1}, \ldots, \mathcal{A}_{k T(k)})$ of $g_{k}$ are associated with parameters $(m_{k1}, \ldots, m_{kT(k)})$, such that $g_{k}(w) = m_{kt}$ for all $w \in \mathcal{A}_{kt}$.

Each regression tree $g_{k}$ is assigned an independent prior as described in \citet{Chipman10}. The distribution on the splitting variable assignments at each interior node and the distribution on the splitting rule assignment in each interior node conditional on the splitting variable are chosen to be uniform. With the default specification by \citet{Chipman10}, there is a 0.83 prior probability that a regression tree only has 2 or 3 terminal nodes. Thus, similar to boosting, the BART prior heavily favours small trees.

The variance parameter $\sigma$ and the node parameters $m_{kt}$ are assigned weakly informative conjugate priors. This enables a convenient and efficient Gibbs sampling algorithm, which \citet{Chipman10} call ``Bayesian backfitting''. Each tree is updated in turn by fitting the partial residuals left over by the other trees. This iterative process of incrementally improving the model fit provides another connection to boosting.

We assume that the observational and experimental data are generated i.i.d. from a pair of true distributions $P_{0}(\cdot \mid F_{T} = \mathcal{O})$ and $P_{0}(\cdot \mid F_{T} = \mathcal{E})$ respectively. The propensity score $\pi(x) = P_{0}(T = 1 \mid x, F_{T} = \mathcal{O})$ is the conditional probability of receiving treatment given the pre-treatment covariates in the observational regime.

As a flexible nonparametric method, BART employs regularization through its prior to achieve good predictive performance. Unfortunately, this regularization can induce a non-negligible bias that bleeds into the estimation of low-dimensional parameters like average treatment effects. The magnitude of the bias can be reduced by ``undersmoothing'' the prior, but this is difficult to implement in practice and is likely to compromise the overall performance of the method.

This type of bias arises in Bayesian semiparametric theory through the so-called ``least favorable direction'' in the model space that contains the minimal amount of information about the parameter. Under mild regularity conditions, it is known that the likelihood function will asymptotically concentrate along the least favorable direction; a heuristic discussion of this phenomenon can be found in \citet{Bickel12}. Intuitively, a key condition for avoiding bias is to have a prior that is approximately invariant in the least favorable direction \citep{Castillo12, Castillo15}. 

For the average treatment effect, the least favorable direction for the outcome regression function is the ``clever covariate''
\begin{equation*}
    h(T,X) = \frac{T}{\pi(X)}- \frac{1-T}{1-\pi(X)}.
\end{equation*}
To help ensure prior invariance, we propose fitting BART on the observational data with the augmented set of covariates $W = (T,X, h(T,X))$. If $\pi$ is unknown, then it is replaced by an estimate $\hat{\pi}$.

The idea of incorporating the clever covariate into a regression model has a long history \citep{Firth98, Scharfstein99} and it is a prominent component of Targeted Maximum Likelihood Estimation (TMLE) \citep{vanderLaan11}. In Bayesian inference, \citet{Ray19} and \citet{Ray20} proposed incorporating an estimate of $h(T,X)$ as a linear covariate to augment a Gaussian process prior. By including the clever covariate as a splitting variable rather than as a separate linear covariate, we allow for more complex relationships between $Y$ and $h(T,X)$ if it is favored by the data; this can give us improved finite sample performance while still offering protection against asymptotic bias. \citet{Hahn20} included an estimate of the propensity score as a splitting variable but was motivated solely by improved estimation of the regression function.

Our augmented BART posterior induces a sequence of predictives $P_{m}(Y \mid T,X)$ for $m \geq n$. For the marginal covariate model, we recommend specifying the Bayesian bootstrap as described in \S\ref{sec::pred_spec}, which is convenient and avoids the type of regularization bias described earlier \citep{Ray19, Ray20}. Since the unobserved target trial data are exchangeable in this case, we can sample $P_{\infty}$ directly from the posteriors rather than implement predictive resampling.

If $\pi$ is unknown and must be estimated, then our method lies outside of a conventional Bayesian framework. This is because the treatment variables and covariates in the observational data are not modeled as being exchangeable. In particular, the \textit{coherence} property \citep{Bissiri16} is violated; the posterior is no longer invariant to the order of updating. However, our method is well-motivated in the predictive resampling framework. We are simply using the augmented BART model to produce a prior for the unobserved target trial data, which induces a sequence of predictives satisfying the martingale condition.

The BART model in (\ref{eqn::cont_out}) is for continuous outcomes, but BART can also be applied more generally by using link functions; see the summary in \citet{Hill20}. Existing semiparametric theory \citep[e.g.][]{vanderLaan11, Ray20} suggests that our augmentation should be used with canonical link functions (e.g. the logistic link rather than the probit link). We leave the formal theoretical development of our model to future work.

\subsection{Marginal prediction with inverse probability weighting} \label{sec::marg_pred}

Traditional Bayesian inference requires a full joint model for the observed and unobserved data. Consequently, a method like inverse probability weighting, which is often viewed as a more robust option to outcome regression modeling when the propensity score is known \citep{RobinsRitov97}, cannot be motivated in a Bayesian framework. So far, we have proceeded by specifying a full joint predictive for the observational regime and transporting it into the experimental regime using the assumptions developed in \S\ref{sec::setup}. In this subsection, we propose a different approach that targets the marginal predictive of $Y$ directly and incorporates inverse probability weighting in a very natural way.

As in \S\ref{sec::clev_cov}, we assume that the observational and experimental data are generated i.i.d. from a pair of true distributions $P_{0}(\cdot \mid F_{T} = \mathcal{O})$ and $P_{0}(\cdot \mid F_{T} = \mathcal{E})$ respectively. We will only consider estimation of the average treatment effect, so it is sufficient for the kernel $P_{0}$ to satisfy Assumptions \ref{ass::pos}-\ref{ass::modularity}, rather than the additional assumptions needed when considering $T^{\mathcal{O}}$. The propensity score $\pi(x) = P_{0}(T = 1 \mid x, F_{T} = \mathcal{O})$ is assumed to be known. 

A simple inverse probability weighted estimator of the average treatment effect is the H\'ajek estimator \citep{Hajek71} 
\begin{equation*}
    \hat{\theta}_{HJ} = \sum_{i=1}^{n}(\lambda_{1i}-\lambda_{0i})Y_{i},
\end{equation*}
where
\begin{align*}
    \lambda_{1i} &=\left(\sum_{j=1}^{n}\frac{T_{j}}{\pi(X_{j})}\right)^{-1}\frac{T_{i}}{\pi(X_{i})} \\
    \lambda_{0i} &=\left(\sum_{j=1}^{n}\frac{1-T_{j}}{1-\pi(X_{j})}\right)^{-1}\frac{1-T_{i}}{1-\pi(X_{i})}.
\end{align*}
We highlight the fact that $\sum_{i=1}^{n}\lambda_{1i} = \sum_{i=1}^{n}\lambda_{0i} = 1$. This motivates our specification of the first-step marginal predictives
\begin{align*}
    P_{n}(Y \mid T_{n+1} = 1, F_{T} = \mathcal{E}) &\sim \sum_{i=1}^{n}\lambda_{1i}\delta_{Y_{i}} \\ 
    P_{n}(Y \mid T_{n+1} = 0, F_{T} = \mathcal{E}) &\sim \sum_{i=1}^{n}\lambda_{0i}\delta_{Y_{i}}, 
\end{align*}
where we have used the H\'ajek weights as predictive probabilities.

The first-step predictives above have an intuitive justification. Suppose that $\pi$ were a constant function. It would be natural to predict $Y_{n+1}$ as follows: predict $T_{n+1}$ by sampling from $\text{Ber}(0.5)$ and then resample from $\{Y_{i}: T_{i} = T_{n+1}, i= 1, \ldots, n\}$, much like the Bayesian bootstrap. On the other hand, if $\pi$ is not constant, the predictives adjust to the experimental regime by putting more weight on the outcomes with smaller $\pi$ and vice-versa.

Our update follows a P\'olya urn scheme; that is, the new predictive is a weighted average of the old predictive and a point mass on the newly predicted outcome value. For $m \geq 1$, the predictives are defined by
\begin{align} \label{eqn::margpred_up1}
    P_{n+m}(Y \mid T_{n+m+1} = 1, F_{T} = \mathcal{E}) &\sim \frac{\Tilde{n}_{1}\sum_{i=1}^{n}\lambda_{1i}\delta_{Y_{i}}+ \sum_{j=1}^{m}T_{n+j}\delta_{Y_{n+j}}}{\Tilde{n}_{1}+\sum_{j=1}^{m}T_{n+j}}, \\ \label{eqn::margpred_up2}
    P_{n+m}(Y \mid T_{n+m+1} = 0, F_{T} = \mathcal{E}) &\sim \frac{\Tilde{n}_{0}\sum_{i=1}^{n}\lambda_{0i}\delta_{Y_{i}}+ \sum_{j=1}^{m}(1-T_{n+j})\delta_{Y_{n+j}}}{\Tilde{n}_{0}+\sum_{j=1}^{m}(1-T_{n+j})},
\end{align}
where $\Tilde{n}_{1}$ and $\Tilde{n}_{0}$ are user-specified effective sample sizes that determine the influence of the observed data relative to the new predictions.

A simple choice for the effective sample sizes is to use the number of individuals with $T_{i} = t$ among the observed data
\begin{equation} \label{eqn::steph_ess}
    \Tilde{n}_{t} = \sum_{i=1}^{n}1(T_{i} = t).
\end{equation}
Alternatively, we could view the H\'ajek weights from an importance sampling perspective, which motivates
\begin{equation} \label{eqn::andr_ess}
    \Tilde{n}_{t} = \frac{1}{\sum_{i=1}^{n}\lambda_{ti}^{2}}.
\end{equation}
from importance sampling methodology \citep[\S4.4 of][]{Robert10}. If $\pi$ is constant in the observed sample, (\ref{eqn::steph_ess}) and (\ref{eqn::andr_ess}) are equal. Otherwise, variability in the H\'ajek weights will reduce the effective sample size in (\ref{eqn::andr_ess}), while (\ref{eqn::steph_ess}) is unchanged. In Theorem \ref{theo::marg_cons}, we will show that both choices lead to the optimal posterior contraction rate for estimating the average treatment effect.

Since (\ref{eqn::margpred_up1}) and (\ref{eqn::margpred_up2}) are Dirichlet-multinomial updates, it is immediate that the predicted outcomes are exchangeable. In fact, the martingale posterior can be accessed directly:
\begin{equation*}
    P_{\infty}(Y \mid T=t, F_{T} = \mathcal{E}) = \sum_{i=1}^{n}\omega_{ti}\delta_{Y_{i}}
\end{equation*}
and
\begin{equation*}
    \theta_{\infty} = E_{\infty}[Y \mid T = 1, F_{T} = \mathcal{E}] - E_{\infty}[Y \mid T = 0, F_{T} = \mathcal{E}] = \sum_{i=1}^{n}(\omega_{1i}-\omega_{0i})Y_{i},
\end{equation*}
where
\begin{equation*}
    (\omega_{t1}, \ldots, \omega_{tn}) \sim \text{Dir}(\Tilde{n}_{t}\lambda_{t1}, \ldots, \Tilde{n}_{t}\lambda_{tn}).
\end{equation*}
This posterior has mean equal to the H\'ajek estimator. 

The following theorem shows that the posterior of $\theta_{\infty}$ contracts at rate $1/\sqrt{n}$ to the truth
\begin{equation*}
    \theta_{0} = E_{P_0}[Y \mid T=1, F_{T} = \mathcal{E}]-E_{P_0}[Y \mid T=0, F_{T} = \mathcal{E}].
\end{equation*}
We use the shorthand notation $Z^{(n)} = (Z_{1},\ldots, Z_{n})$.

\begin{theorem} \label{theo::marg_cons}
Suppose that there exists $\varepsilon > 0$ such that with probability 1, $0 < \varepsilon < P_{0}(T = 1 \mid X, F_{T} =  \mathcal{O}) < 1- \varepsilon < 1$. And suppose also that $\text{var}_{P_0}(Y \mid T=t, F_{T} = \mathcal{E}) < \infty$ for $t \in \{0,1\}$. For either (\ref{eqn::steph_ess}) or (\ref{eqn::andr_ess}) as the choice of effective sample sizes, the posterior of $\theta_{\infty}$ contracts to $\theta_{0}$ at rate $1/\sqrt{n}$; that is, for every $M_{n} \rightarrow \infty$, $\Pi(\theta_{\infty}: \lVert \theta_{\infty}-\theta_{0}\rVert_{2}\geq M_{n}/\sqrt{n} \mid Z^{(n)}) \rightarrow 0$ in $P_{0}(\cdot \mid F_{T} = \mathcal{O})$-probability.
\end{theorem}

Our approach also enables the user to incorporate prior shrinkage into the predictives. For a base probability measure $\Tilde{P}$ and concentration parameter $\alpha > 0$, we can specify
\begin{align*}
    P_{n}(Y \mid T_{n+1} = 1, F_{T} = \mathcal{E}) &\sim \frac{\Tilde{n}_{1}\sum_{i=1}^{n}\lambda_{1i}\delta_{Y_{i}} + \alpha \Tilde{P}}{\Tilde{n}_{1} + \alpha} \\ 
    P_{n+m}(Y \mid T_{n+m+1} = 1, F_{T} = \mathcal{E}) &\sim \frac{\Tilde{n}_{1}\sum_{i=1}^{n}\lambda_{1i}\delta_{Y_{i}}+ \sum_{j=1}^{m-1}T_{n+j}\delta_{Y_{n+j}}+ \alpha \Tilde{P}}{\Tilde{n}_{1}+\sum_{j=1}^{m-1}T_{n+j} + \alpha},
\end{align*}
for $m \geq 1$, with similar expressions for $T=0$. Motivated by the Dirichlet process $\text{Dir}(\alpha\Tilde{P})$ prior from Bayesian nonparametrics, this has the effect of regularizing the predictives towards $\Tilde{P}$, with $\alpha$ playing the role of the prior effective sample size. 

\section{Illustration} \label{sec::appl}

We illustrate our methodology by performing a target trial emulation to study the causal effect of maternal smoking during pregnancy on birthweight. Our dataset is an excerpt of the data studied in \citet{Almond05} and \citet{Cattaneo10}; the same excerpt was also analysed in \citet{Lee17} and is publicly available on the STATA website (http://www.stata-press.com/data/r13/cattaneo2.dta). The dataset contains information on singleton births in Pennsylvania between 1989 and 1991. This particular state and time period was chosen by \citet{Almond05} because complete smoking information was available for over 95 percent of the units.

The treatment policy of interest is smoking cessation during pregnancy; we define the treatment variable $T$ to take the value 1 if the individual is assigned to the smoking cessation group, and 0 if they are instructed to continue smoking. For obvious ethical reasons, it would be impossible to implement the control arm in a real randomized trial. Thus, the best alternative option is to emulate a target trial using observational data. The protocol descriptions for our target trial specification and emulation can be found in Table \ref{tab::app_protocol}.

\begin{table}[h!]
\caption{Target trial protocol components for studying the effect of smoking during pregnancy on birthweight using an excerpt of the \citet{Almond05} dataset. \label{tab::app_protocol}}
\begin{center}
\resizebox{1\textwidth}{!}{\begin{tabular}{|l |p{0.3\textwidth}p{0.2\textwidth}|p{0.3\textwidth}c|} \hline
\multirow{2}{4em}{Protocol component} & \multicolumn{2}{c||}{Target trial protocol} & \multicolumn{2}{c|}{Emulation using observational HIV cohorts}\\ \cline{2-5}
 & \multicolumn{1}{c|}{Description} & \multicolumn{1}{c||}{Generative model} & \multicolumn{1}{c|}{Description} & \multicolumn{1}{c|}{Generative model} \\ \hline
Eligibility criteria  & \multicolumn{1}{p{0.3\textwidth}|}{White and non-Hispanic female smokers scheduled to give birth in Pennsylvannia between 1989 and 1991.

\textbf{Subgroup:} Mothers who gave birth to a singleton child between 1989 and 1991.} & \multicolumn{1}{c||}{$X \in {\cal{X}}$}  & \multicolumn{1}{p{0.3\textwidth}|}{Same as for target protocol, except that non-smokers are also eligible.}& $X \in {\cal{X}}$ \\ \hline
Treatment strategies & \multicolumn{1}{p{0.3\textwidth}|}{Those assigned to the smoking cessation group must not smoke during pregnancy. Those assigned to the no smoking cessation group are instructed to continue smoking.} & \multicolumn{1}{c||}{$T$} & \multicolumn{1}{p{0.3\textwidth}|}{We compare the smokers and non-smokers during pregnancy. This requires the assumption that smokers who stop during pregnancy are comparable with those who never smoked in the context of birthweights.} & $T$ \\ \hline
Treatment assignment & \multicolumn{1}{p{0.3\textwidth}|}{Eligible individual will be randomly assigned to either the smoking cessation group or the no smoking cessation group and will be aware of which strategy they were assigned to. Treatment is assigned at pregnancy discovery.} & \multicolumn{1}{c||}{$P(T \mid F_{T} = \mathcal{E})$} & \multicolumn{1}{p{0.3\textwidth}|}{Emulate randomization by adjusting for the mother's age, educational attainment, marital status, and foreign-born status; the father's age and educational attainment; whether the newborn is the mother's first child, and whether there was a previous birth where the newborn died.} & $P(T \mid X, F_{T} = \mathcal{O})$\\ \hline
Outcome & \multicolumn{1}{p{0.3\textwidth}|}{Child birthweight} & \multicolumn{1}{c||}{$Y$} & \multicolumn{1}{p{0.3\textwidth}|}{Same as for target protocol.} & $Y$\\ \hline
Follow-up & \multicolumn{1}{p{0.3\textwidth}|}{From treatment assignment until childbirth.} & \multicolumn{1}{c||}{} & \multicolumn{1}{p{0.3\textwidth}|}{Same as for target protocol. } & \\ \hline
Causal estimand & \multicolumn{1}{p{0.3\textwidth}|}{Per-protocol effect: the effect of smoking cessation on birthweight for both the whole population and within values of the mother's age.} & \multicolumn{1}{p{0.3\textwidth}||}{Population average treatment effect: $E[Y \mid T=1,F_{T} = \mathcal{E}] - E[Y \mid T=0,F_{T} = \mathcal{E}]$.

Conditional average treatment effect: 

$E[Y \mid T=1, X_{\text{age}},F_{T} = \mathcal{E}] - E[Y \mid T=0,X_{\text{age}},F_{T} = \mathcal{E}]$} & \multicolumn{1}{p{0.3\textwidth}|}{Observational analogue of the per-protocol effect.} & Same as for target protocol.\\ \hline
\end{tabular}}
\end{center}
\end{table}

Following \citet{Lee17}, we restrict our attention to white and non-Hispanic mothers. In our target trial, the eligible individuals are white and non-Hispanic female smokers whom are due to give birth in Pennsylvania between 1989 and 1991. To match the information available in the observational dataset, we further restrict attention to causal effects for the subpopulation whom gave birth to a singleton child. Unfortunately, the observational dataset does not contain information on smoking history before pregnancy, so the trial emulation eligibility criteria will allow for non-smokers. In order to proceed, we must make the assumption that smoking history prior to pregnancy has no effect on the child birthweight, such that smokers who stop during pregnancy are comparable with non-smokers.

For the target trial, treatment is assigned completely at random to one of the two treatment groups at the time of pregnancy discovery. To emulate randomization with the observational data, we adjust for the following pre-treatment covariates, which we denote by $X$: the mother's age, educational attainment, marital status, and foreign-born status; the father's age, educational attainment; whether the newborn is the mother's first child, and whether there was a previous birth where the newborn died.

This example illustrates the utility of target trial emulation to clarify the limitations of the data and the assumptions required to obtain causal inference. We have been explicit that our interest is in smoking cessation, rather than simply comparing smokers and non-smokers, which elucidates the particular intervention for policy-making and the shortcomings of not having information on smoking history. The specification of the eligibility criteria also highlights that if one wishes to generalize the findings to different populations---e.g. white and non-Hispanic mothers in the US in 2022---then it may be necessary to adapt the marginal model for the pre-treatment covariates accordingly. This reduces the risk of carelessly extrapolating the results.

We analyse the data with four different methods. Using the joint method, we compare BART with and without the clever covariate augmentation introduced in \S\ref{sec::clev_cov}. In both cases, we use the Bayesian bootstrap for the marginal covariate model and implement the procedures using the default settings in the \texttt{bartMachine} R package. Since the propensity score is unknown, we estimate it using the posterior mean from a probit-link BART model as implemented in \texttt{bartMachine}. Using this estimated propensity score, we also implement the marginal prediction method from \S\ref{sec::marg_pred} with the two effective sample sizes in (\ref{eqn::steph_ess}) and (\ref{eqn::andr_ess}). The units for the treatment effects are given in grams; across all units in the dataset, the average birthweight was 3416g. Code for producing the results in this section can be found at \url{https://github.com/andrew-yiu/Predictive-causal-inference}.

Following \citet{Lee17}, we first investigate the conditional average treatment effects given the mother's age. The plots for BART with and without the clever covariate augmentation are found in Figure \ref{fig::CATE}. As expected, the intervals for the augmented model are slightly wider due to the additional covariate, but the overall behaviour is very similar. In both cases, we observe a general upward trend in the magnitude of the treatment effect as age increases; the effect of smoking cessation is significant across all ages, but the results suggest that the treatment might be more pertinent for older mothers. This agrees with the findings in \citet{Lee17}. 

The posterior densities for the average treatment effect across all the methods are found in Figure \ref{fig::ATE} and the numerical results are contained in Table \ref{tab::post_summ}. The differences between BART with and without the clever covariate are more noticeable here. Not only is the uncertainty higher for the augmented model, but there is also a slight shift in the centering in the direction of the marginal prediction methods. The uncertainty in the marginal prediction method for the importance sampling effective sample size in (\ref{eqn::andr_ess}) is noticeably higher than the uncertainty for (\ref{eqn::steph_ess}); the total effective sample size for (\ref{eqn::andr_ess}) was 3303 compared to the full sample size of 3754 for (\ref{eqn::steph_ess}). This indicates some variability in the estimate of $\pi$.

\begin{table}[h!]
\caption{Posterior summary statistics for the average treatment effect. ``BART'', BART and the Bayesian bootstrap; ``BART-CC'', BART augmented with the clever covariate and the Bayesian bootstrap; ``\text{Marg-Obs}'', marginal prediction with the effective sample size in (\ref{eqn::steph_ess}), `\text{Marg-IS}'', marginal prediction with the effective sample size in (\ref{eqn::andr_ess}). \label{tab::post_summ}}
\begin{center}
\resizebox{.7\textwidth}{!}{\begin{tabular}{|l |cccc|} \hline
Method & Mean & Median & SD & 95\% interval \\ \hline
BART   & 257.98 & 257.78 & 25.64 & $(209.85, 307.88)$ \\
BART-CC  & 262.64 & 262.71 & 27.82 & $(208.26, 317.47)$\\
Marg-Obs & 267.32 & 267.16 & 23.98 & $(220.81, 316.04)$\\
Marg-IS   & 266.49 & 265.63 & 32.05 & $(203.37, 330.22)$\\ \hline
\end{tabular}}
\end{center}
\end{table}

\begin{figure}[]
\begin{center}
\includegraphics[width=5.5in]{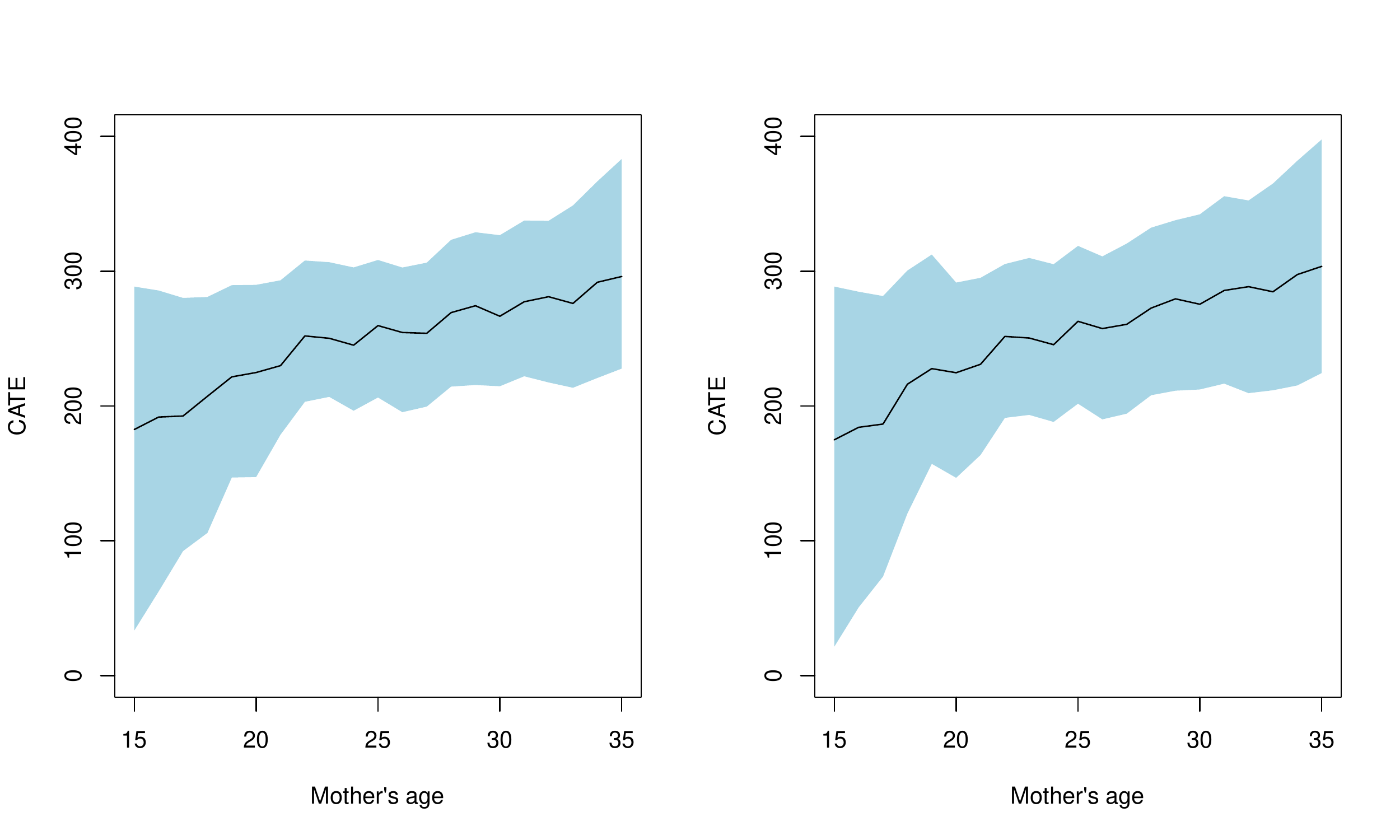}
\end{center}
\caption{The posterior means and pointwise $95\%$ intervals for the conditional average treatment effects given age: BART and the Bayesian bootstrap (left); BART augmented with the clever covariate and the Bayesian bootstrap (right).\label{fig::CATE}}
\end{figure}

\begin{figure}[]
\begin{center}
\includegraphics[width=5.5in]{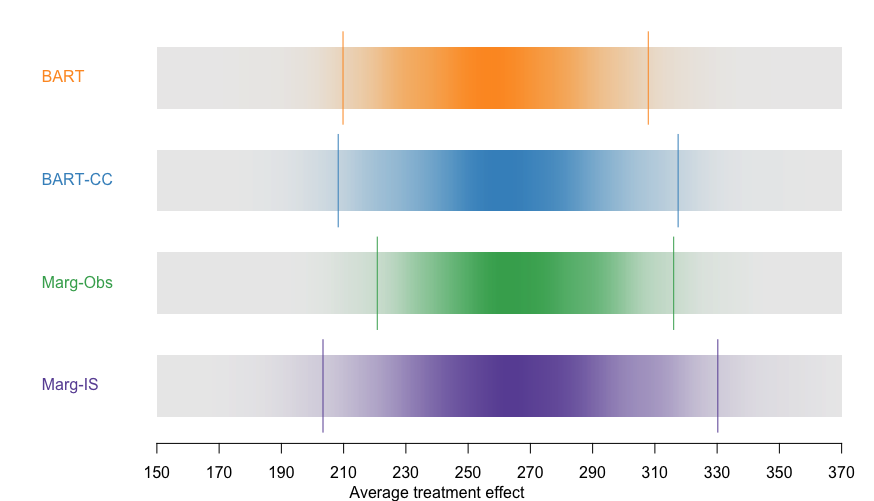}
\end{center}
\caption{Posterior distributions for the average treatment effect. The darkness of the strips is proportional to the posterior density, with the central 95\% credible regions indicated. ``BART'', BART and the Bayesian bootstrap; ``BART-CC'', BART augmented with the clever covariate and the Bayesian bootstrap; ``\text{Marg-Obs}'', marginal prediction with the effective sample size in (\ref{eqn::steph_ess}), `\text{Marg-IS}'', marginal prediction with the effective sample size in (\ref{eqn::andr_ess}). \label{fig::ATE}}
\end{figure}

We have also included the splitting variable inclusion proportions for the augmented BART model in Figure \ref{fig::inc_prop}. These represent the proportion of all splitting rules that use the covariate in the sum-of-trees model averaged across the posterior draws \citep{Chipman10}. For added stability, this is further averaged across 100 independent BART fits. The clever covariate has an inclusion proportion of about 0.07, which places it in the middle of the covariate rankings. This suggests that the clever covariate is a useful transformation of $T$ and $X$.

\begin{figure}[]
\begin{center}
\includegraphics[width=5.5in]{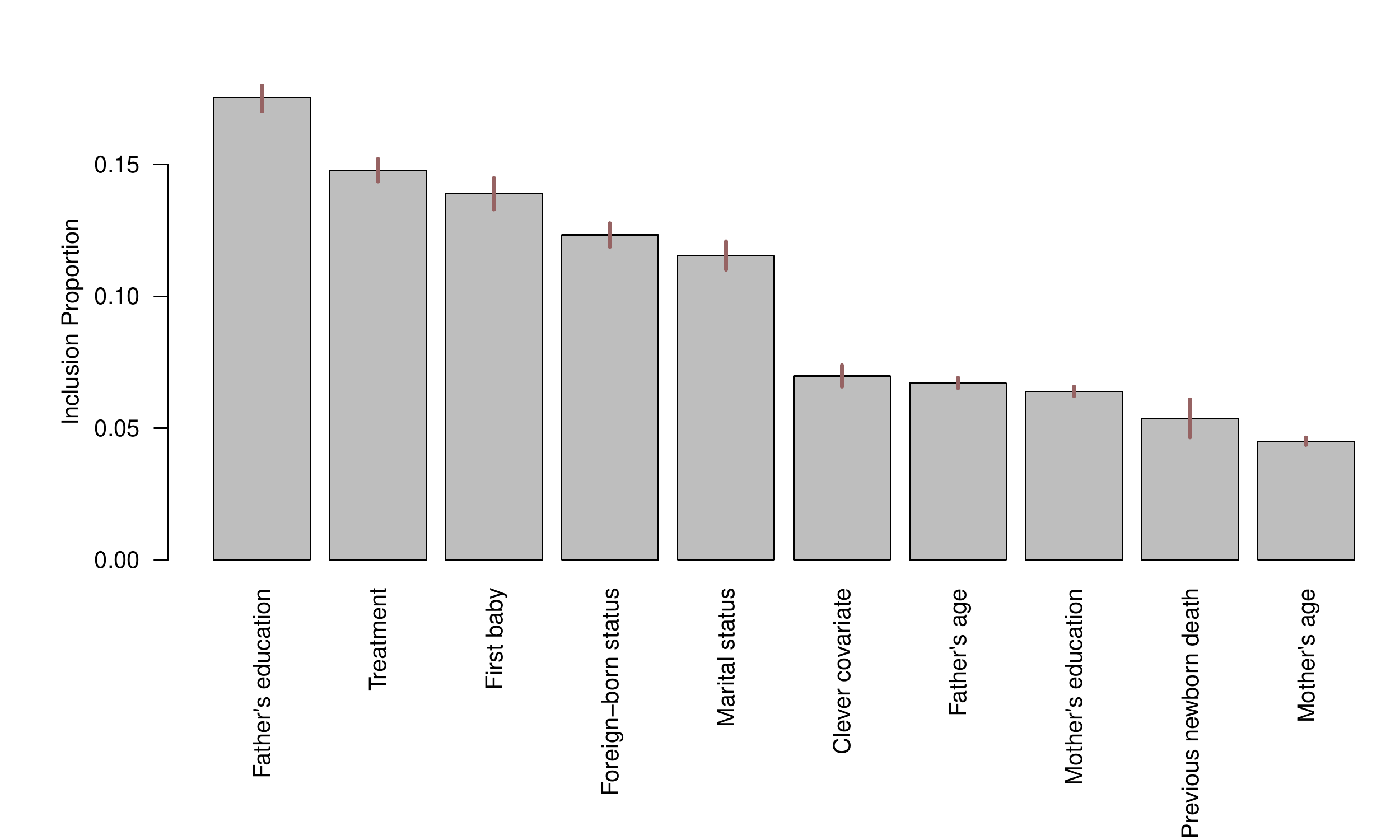}
\end{center}
\caption{Splitting variable inclusion proportions for BART augmented with the clever covariate. The 95\% uncertainty intervals across iterations are indicated by the vertical lines. \label{fig::inc_prop}}
\end{figure}

Overall, the results for all four methods are reasonably similar, with posterior mean estimates of approximately 260g and posterior standard deviations around 30g. The largest difference between the posterior means is 8.51g, which is only around 0.3 posterior standard deviations. Furthermore, our results show concordance with the estimate of 267.55g in \citet{Lee17}, which was derived using a doubly robust estimator with a linear outcome regression model and a logistic regression propensity score model.

As discussed in Section 13.4 of \citet{Hernan20}, the similarities between the methods is reassuring as the ordinary BART method and the marginal prediction methods use different models in orthogonal ways. Thus, unless both models happen to be misspecified to a similar degree in the same direction, the results suggest an absence of serious model misspecification.

\section{Discussion} \label{sec::discussion}

Predictive inference coupled with target trial emulation provides a self-contained statistical framework for causal analysis without the need to introduce counterfactuals. The framework  provides an intuitive definition of causal effects as empirical statistics obtained from a hypothetical population-scale randomized trial. The usual causal assumptions map onto easy to understand criteria regarding the transportability and identification of predictive models where the training and deployment environments differ. 

As discussed in \S\ref{sec::tte}, this framework can be used to facilitate the combining of observational studies with existing experimental data, as the fundamental task remains the same, namely, use all available information to hand to build the best  predictive model for a randomized trial in the remaining population. While we have only considered the setting where only observational data is available, it is straightforward to extend our approach to work with both types of data together. The set-up introduced in \S\ref{sec::setup} provides the language to formalize the assumptions to link the experimental and observational regimes, and then one can build a predictive model conditional on all the information to impute the remaining unobserved target trial data. Additional complexities such as partial-compliance are also handled within the common format of generative models. 

\section*{Acknowledgements}

The authors gratefully acknowledge the following funding sources. \textbf{AY} receives funding from Novo Nordisk. \textbf{CH} is supported by The Alan Turing Institute, the Li Ka Shing Foundation, the Medical Research Council, the EPSRC through the Bayes4Health
grant EP/R018561/1, AI for Science and Government UKRI, and the U.K. Engineering and Physical Sciences Research Council.

\bibliography{paper-ref}

\appendix

\section{Proofs}

\subsection{Proof of Theorem \ref{theo::fact}}

\begin{proof}
We are required to establish the following conditional independences:
\begin{enumerate}
    \item[(I)] $Y \independent (T^{\mathcal{O}}, F_{T}) \mid T,X$
    \item[(II)] $T \independent X \mid T^{\mathcal{O}}, F_{T}$
    \item[(III)] $T \independent (T^{\mathcal{O}},X) \mid F_{T} = \mathcal{E}$
    \item[(IV)] $(T^{\mathcal{O}}, X) \independent F_{T}$.
\end{enumerate}
Condition (I) is implied by combining $Y \independent T^{\mathcal{O}} \mid T,X,F_{T}$ (Assumption \ref{ass::cond_ign} and the definition of $T$ as $T=T^{\mathcal{O}}$ given $F_{T} = \mathcal{O}$) and $Y \independent F_{T} \mid T,X$ (Assumption \ref{ass::ext_mod}(i)) using the contraction property of conditional independence. 

By definition, we recall that $T=T^{\mathcal{O}}$ under $F_{T} = \mathcal{O}$, so we trivially have $
    T \independent X \mid T^{\mathcal{O}}, F_{T} = \mathcal{O}.$
By the weak union property of conditional independence, $T \independent (T^{\mathcal{O}},X) \mid F_{T} = \mathcal{E}$ (Assumption \ref{ass::rand}) implies $T \independent X \mid T^{\mathcal{O}}, F_{T} = \mathcal{O}$. Combining the cases for $F_{T} = \mathcal{O}$ and $F_{T} = \mathcal{E}$ gives condition (II).

Assumption \ref{ass::ext_rand} is (III), and Assumptions \ref{ass::ext_mod}(ii) and \ref{ass::ext_mod}(iii) imply (IV) by the contraction property of conditional independence. 
\end{proof}

\subsection{Proof of Corollary \ref{cor::att_iden}}

\begin{proof}
We have
\begin{align*}
    P(T \mid T=t, T^{\mathcal{O}}=1, F_{T} = \mathcal{E}) &= \frac{\sum_{x}P(Y, T=t, T^{\mathcal{O}}=1,X=x \mid F_{T} = \mathcal{E})}{\sum_{x}P(T=t, T^{\mathcal{O}} = 1, X=x \mid F_{T} = \mathcal{E})}.
\end{align*}
By the formula in (\ref{eqn::gform}), the numerator is equal to
\begin{equation*}
    \sum_{x}P(Y \mid T=t, X=x, F_{T} = \mathcal{O})P(T=t \mid F_{T} = \mathcal{E})P(T^{\mathcal{O}} = 1 \mid X=x,F_{T} = \mathcal{O})P(X=x \mid F_{T} = \mathcal{O}),
\end{equation*}
and the denominator is equal to
\begin{equation*}
    \sum_{x}P(T=t \mid F_{T} = \mathcal{E})P(T^{\mathcal{O}} = 1 \mid X=x,F_{T} = \mathcal{O})P(X=x \mid F_{T} = \mathcal{O}).
\end{equation*}
The $P(T=t \mid F_{T} = \mathcal{E})$ factors cancel, and the result follows from Bayes' theorem.
\end{proof}

\subsection{Proof of Theorem \ref{theo::pred_comp}}

\begin{proof}
All the equalities below are to be understood in the almost sure sense. For $k \geq n+1$,
\begin{align*}
    E[P_{k}(Y,T,T^{\mathcal{O}}, X) \mid \mathcal{G}_{k-1}] &= E[P_{k}(Y \mid T,X)P_{k}(T)P_{k}(T^{\mathcal{O}} \mid X)P_{k}(X)\mid \mathcal{G}_{k-1}] \\
    &= E[P_{k}(Y \mid T,X)P_{k}(T^{\mathcal{O}} \mid X)P_{k}(X)\mid \mathcal{G}_{k-1}]P_{k-1}(T) \\ 
    &= E[E[P_{k}(Y \mid T,X) \mid T_{k},X_{k}, \mathcal{G}_{k-1}]P_{k}(T^{\mathcal{O}} \mid X)P_{k}(X) \mid \mathcal{G}_{k-1}]P_{k-1}(T).
\end{align*}
The first equality follows from the factorization in (\ref{eqn::pred_fact}), the second equality takes the non-stochastic $P_{k}(T)=P_{k-1}(T)$ out of the expectation, and the third equality uses the tower law to condition further on $(T_{k},T^{\mathcal{O}}_{k},X_{k})$. Since $P_{k}(T^{\mathcal{O}} \mid X)$ and $P_{k}(X)$ are $\sigma(T_{k},T^{\mathcal{O}}_{k},X_{k}, \mathcal{G}_{k-1})$-measurable, we can take them out of the inner expectation. 

Now consider the inner expectation
\begin{align*}
    E[P_{k}(Y \mid T,X) \mid T_{k},T^{\mathcal{O}}_{k},X_{k}, \mathcal{G}_{k-1}] &= E[P_{k}(Y \mid T,X) \mid T_{k},X_{k}, \mathcal{G}_{k-1}] \\
    &= P_{k-1}(Y \mid T,X).
\end{align*}
The first equality uses the assumption that $P_{k}(Y \mid T,X)$ is $\sigma(Y_{n+1:k}, T_{n+1:k},X_{n+1:k})$-measurable and the conditional ignorability property $P_{k-1}(Y_{k} \mid T_{k},T^{\mathcal{O}}_{k}, X_{k}) = P_{k-1}(Y_{k} \mid T_{k},X_{k})$. The last term is $\mathcal{G}_{k-1}$-measurable, so we can take it outside of the expectation.

It remains to consider
\begin{align*}
    E[P_{k}(T^{\mathcal{O}} \mid X)P_{k}(X) \mid \mathcal{G}_{k-1}] &= E[E[P_{k}(T^{\mathcal{O}} \mid X) \mid X_{k}, \mathcal{G}_{k-1}]P_{k}(X) \mid \mathcal{G}_{k-1}]\\
    &= E[P_{k}(X) \mid \mathcal{G}_{k-1}]P_{k-1}(T^{\mathcal{O}} \mid X) \\
    &= P_{k-1}(T^{\mathcal{O}} \mid X)P_{k-1}(X).
\end{align*}
Similar to before, the first equality uses the tower law to condition further on $X_{k}$, the second equality uses the assumption to evaluate the inner expectation, which is $\mathcal{G}_{k-1}$-measurable, and the third equality follows by assumption. 
\end{proof}

\subsection{Proof of Theorem \ref{theo::marg_cons}}

\begin{proof}

Let $\mu^{(1)} = E_{\infty}[Y \mid T=1, F_{T} = \mathcal{E}] = \sum_{i=1}^{n}\omega_{1i}Y_{i}$, and let $\mu^{(1)}_{0} = E_{P_0}[Y \mid T=1, F_{T} = \mathcal{E}]$. By Lemma 8.2 of \citet{Ghosal17}, it is sufficient to show that
\begin{align} \label{eqn::cont_eqn1}
    E[\mu^{(1)} \mid Z^{(n)}] &= \mu^{(1)}_{0} + O_{P_0}(1/\sqrt{n}) \\ \label{eqn::cont_eqn2} 
    \text{var}[\mu^{(1)} \mid Z^{(n)}] &= O_{P_0}(1/n)
\end{align}
in order to deduce that the posterior for $\mu^{(1)}$ contracts to $\mu^{(1)}_{0}$ at rate $1/\sqrt{n}$.

We start by establishing (\ref{eqn::cont_eqn1}). Recall that the posterior mean $E[\mu^{(1)} \mid Z^{(n)}]$ is the H\'ajek estimator for the treated individuals, which solves the estimating equation
\begin{equation*}
    \sum_{i=1}^{n}\frac{T_{i}(Y_{i} - E[\mu^{(1)} \mid Z^{(n)}])}{\pi(X_{i})} = 0.
\end{equation*}
Through a simple rearrangement, we deduce that
\begin{equation} \label{eqn::rearrange}
    \sqrt{n}(E[\mu^{(1)} \mid Z^{(n)}] - \mu^{(1)}_{0}) = \frac{\frac{1}{\sqrt{n}}\sum_{i=1}^{n}\frac{T_{i}(Y_{i}-\mu^{(1)}_{0})}{\pi(X_{i})}}{\frac{1}{n}\sum_{i=1}^{n} \frac{T_{i}}{\pi(X_{i})}}.
\end{equation}
For the summands in the numerator, we have
\begin{align*}
    E_{P_0}\left[\frac{T(Y-\mu^{(1)}_{0})}{\pi(X)} \mid F_{T} = \mathcal{O}\right] &= 0 \\
    \text{var}_{P_0}\left[\frac{T(Y-\mu^{(1)}_{0})}{\pi(X)} \mid F_{T} = \mathcal{O}\right] &= E_{P_0}\left[\left(\frac{T(Y-\mu^{(1)}_{0})}{\pi(X)}\right)^{2} \mid F_{T} = \mathcal{O}\right] \\
    &= E_{P_0}\left[\frac{E_{P_0}[(Y-\mu^{(1)}_{0})^{2} \mid T=1,X, F_{T} = \mathcal{O}]}{\pi(X)} \mid F_{T} = \mathcal{O}\right] \\
    &= E_{P_0}\left[\frac{E_{P_0}[(Y-\mu^{(1)}_{0})^{2} \mid T=1,X, F_{T} = \mathcal{E}]}{\pi(X)} \mid F_{T} = \mathcal{E}\right] \\
    &< \frac{1}{\varepsilon}\text{var}_{P_0}[Y \mid T=1, F_{T} = \mathcal{E}].
\end{align*}
The last line above is finite by assumption. Thus, by the central limit theorem, the numerator in (\ref{eqn::rearrange}) converges in distribution to a normal distribution with finite variance. For the denominator of (\ref{eqn::rearrange}), the weak law of large numbers implies
\begin{equation} \label{eqn::weaklaw1}
    \frac{1}{n}\sum_{i=1}^{n}\frac{T_{i}}{\pi(X_{i})} \xrightarrow[]{P_{0}} 1.
\end{equation}
Combining the results using Slutsky's lemma establishes (\ref{eqn::cont_eqn1}).

Using Corollary 3.4 of \citet{Kuo18}, the posterior variance of $\mu_{1}$ is
\begin{equation} \label{eqn::post_var}
    \text{var}[\mu^{(1)} \mid Z^{(n)}] = \frac{1}{\Tilde{n}_{1}+1}\left[\frac{\sum_{i=1}^{n}\frac{T_{i}Y^{2}_{i}}{\pi(X_{i})}}{\sum_{j=1}^{n}\frac{T_{j}}{\pi(X_{j})}} -\left\{\frac{\sum_{i=1}^{n}\frac{T_{i}Y_{i}}{\pi(X_{i})}}{\sum_{j=1}^{n} \frac{T_{j}}{\pi(X_{j})}}\right\}^{2}\right].
\end{equation}
The expression in the curly brackets is the posterior mean for the treated $E[\mu^{(1)} \mid Z^{(n)}]$, which converges in $P_{0}$-probability to the truth $E_{P_0}[Y \mid T=1, F_{T} = \mathcal{E}]$ from earlier. Furthermore, we have
\begin{equation} \label{eqn::weaklaw2}
    \frac{1}{n}\sum_{i=1}^{n}\frac{T_{i}Y^{2}_{i}}{\pi(X_{i})} \xrightarrow[]{P_{0}} E_{P_0}[Y^{2}\mid T=1, F_{T} = \mathcal{E}].
\end{equation}
by the weak law of large numbers, where the right-hand side is finite by assumption. Combining (\ref{eqn::weaklaw1}) and (\ref{eqn::weaklaw2}) with Slutsky's lemma, the term in square brackets on the right of (\ref{eqn::post_var}) converges to $\text{var}_{P_0}[Y \mid T=1, F_{T} = \mathcal{E}]$, which is finite by assumption.  

First consider (\ref{eqn::steph_ess}) as the choice of effective sample sizes. By the weak law of large numbers, we have
\begin{equation*}
    \frac{\Tilde{n}_{1}}{n}\xrightarrow[]{P_{0}} P_{0}(T = 1).
\end{equation*}
This establishes (\ref{eqn::cont_eqn2}) by Slutsky's lemma. 

With (\ref{eqn::andr_ess}), we have
\begin{equation*}
    \frac{\Tilde{n}_{1}}{n} = \frac{\left(\sum_{i=1}^{n}\frac{T_{i}}{\pi(X_{i})}\right)^{2}}{\left(\sum_{j=1}^{n}\frac{T_{j}^{2}}{n\pi(X_{j})^{2}}\right)} = \frac{\left(\frac{1}{n}\sum_{i=1}^{n}\frac{T_{i}}{\pi(X_{i})}\right)^{2}}{\left(\frac{1}{n}\sum_{j=1}^{n}\frac{T_{j}^{2}}{\pi(X_{j})^{2}}\right)}.
\end{equation*}
By the weak law of large numbers, 
\begin{align*}
    \left(\frac{1}{n}\sum_{i=1}^{n}\frac{T_{i}}{\pi(X_{i})}\right)^{2} &\xrightarrow[]{P_{0}} 1\\
    \frac{1}{n}\sum_{j=1}^{n}\frac{T_{j}^{2}}{\pi(X_{j})^{2}} &\xrightarrow[]{P_{0}} E_{P_0}\left[\frac{1}{\pi(X)}\right],
\end{align*}
so
\begin{equation*}
    \frac{\Tilde{n}_{1}}{n}\xrightarrow[]{P_{0}} E_{P_0}\left[\frac{1}{\pi(X)}\right]^{-1}.
\end{equation*}

For both choices of effective sample sizes, (\ref{eqn::cont_eqn2}) is established by Slutsky's lemma. A similar argument can be made for $T=0$.
\end{proof}
}

\end{document}